\documentclass[10pt,eqsecnum,twocolumn,pra]{revtex4}
\usepackage[usenames,dvipsnames]{color} 
\usepackage{graphicx}  
\definecolor{LinkColor}{rgb}{0,0,.5} 
 
\newcommand{\ac}[1]{\textsf{\color[rgb]{0.10,0.10,0.60}#1}}    
\usepackage[colorlinks=true,linkcolor=Red,citecolor=LinkColor,urlcolor=LinkColor]{hyperref}

\usepackage{amsthm}
\usepackage{amsmath}
\usepackage{amssymb}
\usepackage{bbm,amsfonts}

\theoremstyle{definition}

\newcommand{\Bsim}{\text{B}_{\text{sim}}}
\newcommand{\Brec}{\text{B}_{\text{rec}}}

\newcommand{\Wal}{w}

\newtheorem{theorem}{Theorem}

\newtheorem{definition}{Definition}

\theoremstyle{definition}

\def\>{\rangle}
\def\<{\langle}

\begin{document}

\title{Compressing Measurements in Quantum Dynamic Parameter Estimation}

\author{Easwar Magesan}
\affiliation{Nuclear Science and Engineering Department and Research Laboratory of Electronics, Massachusetts Institute of Technology, Cambridge, MA 02139, U.S.A.}	
\author{Alexandre Cooper}
\affiliation{Nuclear Science and Engineering Department and Research Laboratory of Electronics, Massachusetts Institute of Technology, Cambridge, MA 02139, U.S.A.}	
\author{Paola Cappellaro}
\affiliation{Nuclear Science and Engineering Department and Research Laboratory of Electronics, Massachusetts Institute of Technology, Cambridge, MA 02139, U.S.A.}

\begin{abstract}
We present methods that can provide an exponential savings in the resources required to perform dynamic parameter estimation using quantum systems. The key idea is to merge classical compressive sensing techniques with quantum control methods to efficiently estimate time-dependent parameters in the system Hamiltonian. We show that incoherent measurement bases and, more generally, suitable random measurement matrices  can be created by performing simple control sequences on the quantum system. Since random measurement matrices satisfying the restricted isometry property  can be used to reconstruct \emph{any} sparse signal in an efficient manner, and many physical processes are approximately sparse in some basis, these methods can potentially be useful in a variety of applications such as quantum sensing and magnetometry. We illustrate the theoretical results throughout the presentation with various practically relevant numerical examples.
\end{abstract}


\maketitle

\section{Introduction}\label{sec:Introduction}

Quantum sensors have emerged as promising devices to beat the shot-noise limit in metrology and, more broadly, to perform measurements at the nanoscale. 
In particular, quantum systems can be used to perform parameter estimation, where the goal is to estimate a set of unknown parameters by manipulating the system dynamics and measuring the resulting state. 
A typical scheme for parameter estimation can be cast in the form of \emph{Hamiltonian identification}, whereby one couples the quantum system to external degrees of freedom so that the parameter of interest is embedded in the Hamiltonian governing the system evolution. Estimates of the desired parameter can then be obtained by estimating the relevant quantities in the Hamiltonian. 
There can be significant advantages in performing sensing using quantum systems, for instance, gains in terms of sensitivity~\cite{GLM} and both spatial ~\cite{GDT} and field amplitude resolution~\cite{JLS}. In general however, parameter estimation can be a difficult problem, especially when the parameters and Hamiltonian are time-dependent. In addition, quantum measurements  can be time-intensive and thus a costly resource for practical parameter estimation schemes. 
The goal of this paper is to provide methods for performing dynamic parameter estimation in a robust manner, while significantly reducing the number of required measurements for high fidelity estimates.

The general scenario we are interested in is when the Hamiltonian of the system can be written as
\begin{equation}
\hat{H}(t)=\left[\frac{\omega_0}{2}+\gamma b(t)\right]\sigma_z,\label{eq:Hamiltonian}
\end{equation}
where we have set $\hbar=1$ and we are interested in reconstructing  over some time interval $I=[0,T]$ a field $b(t)$, which is coupled to the qubit by an interaction of strength $\gamma$. To make the presentation more concrete, throughout we will consider a specific application of dynamic parameter estimation; quantum magnetometry. Applications of magnetic field sensing with quantum probes can be found in a wide variety of emerging research areas such as biomedical imaging~\cite{SVM}, cognitive neuroscience~\cite{HBT,Pham11}, geomagnetism~\cite{Ale2010}, and detecting gravitational waves~\cite{Bla91}. We emphasize that, although we will use the language of magnetometry throughout the rest of the presentation, the methods we propose can  be applied in generality as long as Eq.~(\ref{eq:Hamiltonian}) describes the evolution of the system, up to redefinition of the constants. In the setting of magnetometry, we assume our quantum sensor is a single spin-$\frac{1}{2}$ qubit operating under the Zeeman effect, an example of which is the NV center in diamond~\cite{MSH}. We expect that many of the methods discussed in the context of a spin qubit could be adapted for other types of quantum sensors, such as superconducting quantum interference devices (SQUID's)~\cite{JLS}, nuclear magnetic resonance (NMR) and imaging (MRI) techniques~\cite{Rabi38,Dam71}, and quantum optical and atomic magnetic field detectors~\cite{Bloom62,DHT}. 

Let $b(t)$ represent the magnetic field of interest; setting the gyromagnetic ratio, $\gamma$, equal to 1 and moving into the rotating frame gives the unitary evolution 
\begin{align}
e^{-i\int_0^tH(t^{\prime})dt^{\prime}} = e^{-i\int_0^tb(t^{\prime})dt^{\prime}}\sigma_z.
\end{align}
If $b(t)$ is constant, $b(t)=b$, one can use a simple Ramsey experiment~\cite{Ram50} to determine $b$. One realization of the Ramsey protocol consists of implementing a $\frac{\pi}{2}$ pulse about $X$, waiting for time $T$, implementing a $-\frac{\pi}{2}$ pulse about $Y$, and performing a measurement in the computational basis $\{|0\rangle, |1\rangle\}$. Repeating these steps many times allows one to gather measurement statistics and estimate $b$ since the probability $p_0$ of obtaining outcome $0$ is
\begin{align}
p_0&=\frac{1+\sin\left(\int_{t=0}^Tb dt\right)}{2} = \frac{1+\sin(bT)}{2}.
\end{align}
Under the assumption that $bT \in \left[-\frac{\pi}{2},\frac{\pi}{2}\right]$, we can unambiguously determine $b$ from the above equation. When $b(t)$ is not constant, one must design a new protocol for reconstructing the profile of $b(t)$ since the usual destructive nature of quantum measurements implies that continuous monitoring is not possible. For example, one could partition $I=[0,T]$ into $n$ uniformly spaced intervals and measure $b(t)$ in each interval. However, this is often impractical since, in order for the discretized reconstruction of $b(t)$ to be accurate, $n$ must be large. This entails large overhead associated with readout and re-initialization of the sensor. Recently, Ref.~\cite{CMYC} proposed an alternative method to estimating time-dependent parameters with quantum sensors. In this paper, we discuss how one can build on that result by merging compressive sensing (CS) techniques~\cite{CRT,Don06} and quantum control methods to reproduce the profile of $b(t)$ with the potential for an exponential savings in the number of required measurements. From a more general perspective, compressive sensing techniques can be an ideal potential solution to the problem of costly measurement resources in quantum systems.

Compressive sensing (CS) is a relatively new sub-field of signal processing that can outperform traditional methods of \textit{transform coding}, where the goal is to acquire, transform, and store signals as efficiently as possible. Suppose the signal $F$ of interest is either naturally discrete or discretized into an element of $\mathbb{R}^n$. When the signal is sparse in some basis $\Psi$ of $\mathbb{R}^n$, most traditional methods rely on measuring the signal with respect to a complete basis $\Phi$, transforming the acquired data into the natural (sparse) basis $\Psi$, and finally compressing the signal in this basis by searching for and keeping only the largest coefficients. The end result is an encoded and compressed version of the acquired signal in its sparse basis that can be used for future communication protocols. There are various undesirable properties of this procedure. First, the number of measurements in the basis $\Phi$ is typically on the order of $n$, which can be extremely large. Second, transforming the signal from the acquisition basis $\Phi$ to its natural basis $\Psi$ can be computationally intensive. Lastly, discarding the small coefficients involves locating the largest ones, which is a difficult search problem. Thus, it is clearly desirable to have methods that minimize the number of required measurements, maximize the information contained in the final representation of $F$ in $\Psi$, and circumvent performing large transformations to move between representations.

CS theory shows that one can design both a ``suitable" measurement matrix $\Phi$ and an efficient convex optimization algorithm so that only a small number of measurements $m \ll n$ are required for \emph{exact} reconstruction of the signal $F$. Hence, the compression is performed at the measurement stage itself and all three of the problems listed above are solved; there is a significant reduction in the number of measurements and, since the convex optimization algorithm directly finds the sparse representation in an efficient manner, the reconstruction is exact and no large basis transformation is required. Finding suitable measurement matrices and reconstruction algorithms are active areas of research in compressive sensing and signal processing theory. CS techniques have already been applied in a wide array of fields that include, but is certainly not limited to, medical imaging~\cite{LDP}, channel and communication theory~\cite{TH08}, computational biology~\cite{DSMB}, geophysics~\cite{LH07}, radar techniques~\cite{BS07}, tomography~\cite{GYFB}, audio and acoustic processing~\cite{GHT11}, and computer graphics~\cite{SD11}. The wide applicability of CS techniques is an indication of both its power and generality, and here we show many of these techniques are also amenable to quantum sensing. In the realm of quantum metrology, CS has been used for Hamiltonian identification in the case of static interactions~\cite{SKMR}, and more generally for quantum process tomography~\cite{GYFB}. In contrast, here we introduce methods for dynamic parameter estimation

The paper is organized as follows. In Sec.~\ref{sec:CS} we provide a three-part review of the relevant CS results we will need throughout the presentation.
The first part discusses compressive sensing from the viewpoint of the sparse and measurement bases being fixed and incoherent, which provides the foundation for understanding CS. The second part discusses how one can use randomness to create measurement bases that allow the reconstruction of \emph{any} sparse signal.  In the final part, we discuss the extent to which CS is robust against both approximate sparseness and noise.

We then move on to the main results of the paper. We first show in Sec.~\ref{sec:Main1} how one can utilize both the discrete nature of quantum control sequences and the existence of the Walsh basis to reconstruct signals that are sparse in the time-domain. As an example application, we consider reconstructing magnetic fields produced by spike trains in neurons.
 We then generalize the discussion to \emph{arbitrary} sparse bases in Sec.~\ref{sec:Main2} and show the true power of using CS in quantum parameter estimation. We show that for any deterministic dynamic parameter that is sparse in a known basis, one can implement randomized control sequences and utilize CS to reconstruct the magnetic field. Hence, since this procedure works for signals that are sparse in any basis, it can be thought of as a ``universal" method for performing dynamic parameter estimation. We also show these protocols are robust in the cases of approximate sparsity and noisy signals. We provide various numerical results to quantify these results 
and make concluding remarks in Sec.~\ref{sec:Conclusion}.

\section{Review of Compressive Sensing Theory}\label{sec:CS}


Compressive sensing techniques~\cite{CRT,Don06} allow for the reconstruction of signals using a much smaller number of non-adaptively chosen measurements than required by traditional methods. The success of CS is based on the fact that signals that are naturally sparse in some basis can be efficiently reconstructed using a small number of measurements \emph{if} the measurement (sensing) basis is \textit{``incoherent"}~\cite{DH01} with the sparse basis. Loosely speaking, incoherence generalizes the relationship between time and frequency, and makes precise the idea that the sensing basis is spread out in the sparse basis (and vice versa). As discussed in the introduction, many situations of interest, such as medical imaging~\cite{LDP} and communication theory~\cite{TH08} deal with sensing sparse signals. 
The main advantages of CS over traditional techniques derive from \emph{the compression of the signal  at the sensing stage} and \emph{the efficient and exact convex reconstruction procedures}. 
Let us now describe the general ideas and concepts of CS theory, highlighting those that will be important for the purpose of dynamic parameter estimation with quantum probes.

Suppose $F \in \mathbb{R}^n$ is the deterministic signal we want to reconstruct and $F$ is $S$-sparse when written in the basis $\Psi:=\{\psi_j\}_{j=1}^n$,
\begin{align}
F&=\sum_{j=1}^n\langle F,\psi_j\rangle \psi_j =  \sum_{j=1}^nF_j^{\Psi}\psi_j,
\end{align}
that is, only $S$\ac{$\leq n$} of the coefficients $F_j^{\psi}$ are non-zero. For simplicity, let us assume that $\Psi$ is ordered so that the magnitude of the coefficients $F_j^{\Psi}$ monotonically decrease as $j$ increases. In most physically realistic scenarios, which include quantum parameter estimation, measurements of $F$ are modeled as linear functionals on $\mathbb{R}^n$. By the Riesz representation theorem~\cite{Rudin91}, each measurement can be associated to a unique element of $\mathbbm{R}^n$. Suppose we have access to a set of $n$ orthonormal measurements represented by the orthonormal basis of $\mathbb{R}^n$, $\Phi:=\{\phi_k\}_{k=1}^n$. Since we can represent $F$ as
\begin{align}
F&=\sum_{j=1}^n\langle F,\phi_j\rangle\phi_j = \sum_{j=1}^nF_j^{\Phi}\phi_j,
\end{align}
the output of a measurement $\phi_k$ is the $k$'th coefficient, $\langle F,\phi_k\rangle$, of $F$ with respect to this basis.
One of the central questions compressive sensing attempts to answer is, under the assumption of $S$-sparsity, do there exist 
\begin{enumerate}
\item conditions on the pair $(\Phi,\Psi)$ and 
\item efficient reconstruction algorithms
\end{enumerate}
that allow one to reconstruct $f$ using a small number of measurements from $\Phi$? Ref.~\cite{CRT} has shown that if $(\Phi,\Psi)$ is an incoherent measurement basis pair then one can use an $l_1$-minimization algorithm to efficiently reconstruct $F$ with a small number of measurements. Let us explicitly define these concepts and present a set of well-known compressive sensing theorems that we have ordered in terms of increasing complexity.


\subsection{Compressive Sensing from Incoherent Bases}\label{sec:CS1}

The initial formulation of compressive sensing~\cite{CRT,Don06} required the natural ($\Psi$) and measurement ($\Phi$) bases to be incoherent, which loosely speaking means that these bases are as far off-axis from each other as possible. Rigorously, coherence is defined as follows~\cite{DH01}.
\begin{definition}
Coherence

\medskip

If $(\Phi,\Psi)$ is a basis pair then the coherence between $\Phi$ and $\Psi$ is defined to be
\begin{equation}
\mu(\Phi,\Psi)=\sqrt{n}\: \text{max}_{1\leq i,j \leq n}|\langle \phi_i,\psi_j\rangle|.
\end{equation}
\end{definition}

\medskip

\noindent Thus, the coherence is a measure of the largest possible overlap between elements of $\Phi$ and $\Psi$. If $\Phi$ and $\Psi$ are orthonormal bases then
\begin{equation}
\mu(\Phi,\Psi) \in \left[1,\sqrt{n}\right].
\end{equation}
When $\mu(\Phi,\Psi)$ is close to $1$, $\Phi$ and $\Psi$ are said to be incoherent and when the coherence is equal to 1, the pair is called maximally incoherent. CS techniques provide significant advantages when $\Phi$ and $\Psi$ are incoherent. Suppose we select $m$ measurements uniformly at random from $\Phi$ and we denote the set of chosen indices by $M \subset \{1,...,n\}$ so that $|M|=m$. The first CS theorem~\cite{CR} gives a bound on the probability that the solution to the convex optimization problem,

\medskip
\noindent
\underline{Convex Optimization Problem (COP~1)} \vspace{4pt}\\ $\textrm{argmin}  \{\|x\|_1 \: : \: x \in \mathbb{R}^n\} \: \: \text{subject to} : \: \: \forall j \in M,\: \: $
\begin{align}
F_j^{\Phi}&=\left\langle\phi_j,\sum_{k=1}^nx_k\psi_k\right\rangle,
\end{align}
is equal to the vector $F^{\Psi}$, where ``argmin" refers to finding the vector $x \in \mathbb{R}^n$ that minimizes the 1-norm $\|x\|_1$.

\begin{theorem}\label{thm:Compressive1}

Let $\Phi$ be the sensing basis and $F \in \mathbb{R}^n$ be $S$-sparse in its natural basis $\Psi$. Then, if $\delta > 0$, and
\begin{equation}
m\geq C \mu(\Phi,\Psi)^2 S \log\left(\frac{n}{\delta}\right)\label{eq:mbound}
\end{equation}
for a fixed constant $C$, we have that the solution to COP~1 is equal to $F^{\Psi}$ with probability no less than $1-\delta$.

\end{theorem}

\medskip

\noindent Note that the constant $C$ in Eq.~(\ref{eq:mbound}) is independent of $n$ and is typically not very large. A general rule of thumb is the ``factor of 4" rule which says that approximately $m=4S$ measurements usually suffice when the bases are maximally incoherent. To summarize, under the conditions of Theorem~\ref{thm:Compressive1}, the vector of coefficients in the $\Psi$ basis that minimizes the 1-norm \emph{and} is consistent with the $m$ measurement results will be equal to $\left(F_1^{\Psi},...,F_n^{\Psi}\right)$ with probability no less than $1-\delta$.

Clearly, there are two important factors in Eq.~(\ref{eq:mbound}), the sparsity $S$ and the level of coherence $\mu$ between the bases $\Psi$ and $\Phi$. When the bases are maximally coherent, there is no improvement over estimating all $n$ coefficients. However, when the bases are maximally incoherent, one needs to only perform $O\left(S\log\left(\frac{n}{\delta}\right)\right)$ measurements, which is a significant reduction in measurements (especially for large $n$). There are various examples of incoherent pairs, for instance, it is straightforward to verify that the pairs
\begin{enumerate}
\item standard basis/Fourier basis,
\item  standard basis/Walsh basis,
\item noiselets/wavelets~\cite{CR},
\end{enumerate}
are incoherent, with the first two being maximally incoherent. The second pair will be especially useful for using CS to perform magnetometry using quantum systems. We now analyze how to relax the incoherence condition by using random matrices.

\subsection{Random Measurement Matrices}\label{sec:CS2}

As we have seen, CS techniques can provide significant advantages when the measurement and sparse bases are incoherent. However, for a given sparse basis, the requirement of incoherence places restrictions on the types of measurements one can perform. To overcome this, a large body of theory has been developed regarding how to construct measurement matrices that still afford the advantages of CS. To generalize the discussion, we can see that the convex optimization problem given in Theorem~\ref{thm:Compressive1} can be put into the following generic form

\medskip
\noindent
\underline{Convex Optimization Problem (COP~2)} \vspace{4pt}\\
$\textrm{argmin}  \{\|\tilde{x}\|_1 \: : \: \tilde{x} \in \mathbb{R}^n\} \: \: \text{subject to} \:\: y=A\tilde{x}$.

\medskip

\noindent This can be seen by taking $A=R\Phi^{\dagger}\Psi$, where $R$ is an $m\times n$ matrix that picks out $m$ rows from $\Phi^{\dagger}\Psi$. Focusing on this form of the convex optimization problem, we first discuss conditions on $A$ which ensure exact recovery of sparse signals before describing how to construct such $A$.

\subsubsection{Restricted Isometry Property}\label{sec:RIP}

The restricted isometry property and constant are defined as follows~\cite{CT05}.
\begin{definition}\label{def:RIP}
Restricted Isometry Property (RIP)
\end{definition}

We say that a matrix $A$ satisfies the restricted isometry property (RIP) of order $(S,\delta)$ if $\delta \in (0,1)$ is such that for all $S$-sparse vectors $x \in \mathbbm{R}^n$, 
\begin{equation}
(1-\delta)\|x\|_2^2 \leq \|Ax\|_2^2 \leq (1+\delta)\|x\|_2^2.
\end{equation}

\medskip

\noindent Note that this is equivalent to:
\begin{enumerate}
\item The spectral radius of $A^TA$, denoted $\sigma(A^TA)$, lying in the range $(1-\delta,1+\delta)$,
\begin{equation}
1-\delta \leq \sigma(A^TA) \leq 1+\delta,
\end{equation}
\item and also
\begin{equation}
\sqrt{1-\delta} \leq \|A\|_{2,S} \leq \sqrt{1+\delta},
\end{equation}
\noindent where we have defined the matrix 2-norm of sparsity level $S$ for $A$,
\begin{equation}
 \|A\|_{2,S} = \max_{S\text{-sparse} \: x\: : \|x\|_2=1}\|Ax\|_2.
\end{equation}
\end{enumerate}

\begin{definition}
Restricted Isometry Constant (RIC)
\end{definition}

The infimum over all $\delta$, denoted $\delta_S$, for which $A$ satisfies the RIP at sparsity level $S$ is called the restricted isometry constant (RIC) of $A$ at sparsity level $S$. We also say that $A$ satisfies the RIP of order $S$ if there is some $\delta \in (0,1)$ for which $A$ satisfies the RIP of order $(S,\delta)$. 

\medskip

The RIP is fundamental in compressive sensing. If $A$ satisfies the RIP with $\delta_S \ll 1$ then $A$ acts like an isometry on $S$-sparse vectors, that is, it preserves the Euclidean 2-norm of $S$-sparse vectors. Hence, $S$-sparse vectors are \emph{guaranteed} to not be in the kernel of $A$ and, if $A$ constitutes a measurement matrix, one might hope $x$ can be reconstructed via sampling from $A$. As shown in Theorem~\ref{thm:CompressiveRIP1} below, which was first proved in~\cite{Foucart2010}, this is indeed the case.

\subsubsection{Reconstruction for Sparse Signals}\label{sec:reconstructionsparse}

\begin{theorem}\label{thm:CompressiveRIP1}

Suppose $x \in \mathbb{R}^n$ satisfies the following conditions
\begin{enumerate}
\item $x$ is $S$-sparse, 
\item the underdetermined $m \times n$ matrix $A$ satisfies the RIP of order $2S$ with $\delta_{2S} < 0.4652$,
\item $y=Ax$. 
\end{enumerate}
Then the there is a unique solution $x^*$ of COP~2, $x^*=x$.
\end{theorem}

\medskip

\noindent The constant $0.4652$ is not known to be optimal. It is important to note that Theorem~\ref{thm:CompressiveRIP1} is not probabilistic. If $A$ satisfies the RIP of order $2S$, and the associated RIC $\delta_{2S}$ is small enough, then exact recovery will always occur. Thus, recalling that Theorem~\ref{thm:Compressive1} was probabilistic, it is clear that even if the basis pair $(\Phi,\Psi)$ is incoherent, the matrix $R\Phi^{\dagger}\Psi$ is not guaranteed to satisfy the RIP property for $m$ given by Eq.~(\ref{eq:mbound}). Equivalently, choosing this many rows at random from $\Phi$ is not guaranteed to produce a matrix $R\Phi^{\dagger}\Psi$ which satisfies the RIP of order $2S$ with $\delta_{2S} < 0.4652$. However, if we allow for $m \sim O\left(S(\log (n))^4\right)$, then $R\Phi^{\dagger}\Psi$ does satisfy the RIP with probability 1. 

With Theorem \ref{thm:CompressiveRIP1} in hand, the remaining question is how to create $m\times n$ matrices $A$ that satisfy the RIP with small $\delta_{2S}$. Deterministically constructing such $A$ is a hard problem, however various random matrix models satisfy this condition with high probability if $m$ is chosen large enough. For a detailed discussion, see Ref.~\cite{Baraniuk2007}. The key result is the following theorem.

\begin{theorem}\label{thm:RIP}
 Suppose
\begin{enumerate}
\item $S$, $n$, and $\delta \in (0,1)$ are fixed and
\item the $mn$ entries of $A$ are chosen uniformly at random from a probability distribution to form a matrix-valued random variable $A(\omega)$ which, for all $\epsilon \in (0,1)$, satisfies the concentration inequality
\begin{align}\label{eq:concentration}
\mathbbm{P}\left[\left| \left\| A(\omega)x\right\|_2^2 - \left\|x\right\|_2^2\right| \geq \epsilon \left\|x\right\|_2^2\right] \leq 2e^{-nC_0(\epsilon)}.
\end{align}
Note that Eq.~(\ref{eq:concentration}) means that the probability of a randomly chosen vector $x$ satisfying $\left| \left\| A(\omega)x\right\|_2^2 - \left\|x\right\|_2^2\right| \geq \epsilon \left\|x\right\|_2^2$ is less than or equal to $2e^{-nC_0(\epsilon)}$, where $C_0(\epsilon) > 0$ is a constant that depends only on $\epsilon$.
\end{enumerate}
Then there exist constants $C_1$, $C_2 > 0$ that depend only on $\delta$ such that, if
\begin{equation}
m\geq C_1S\log\left(\frac{n}{S}\right),
\end{equation}
then, with probability no less than $1-2e^{-C_2n}$, $A$ satisfies the RIP of order $(S,\delta)$.
\end{theorem}

We note some important points of Theorem~\ref{thm:RIP}. First, in the spirit of Sec.~\ref{sec:CS1}, let us fix some \emph{arbitrary} basis $\Psi$ and choose a random $m\times n$ measurement matrix $R\Phi^{\dagger}$ according to a probability distribution that satisfies Eq.~(\ref{eq:concentration}). Then Theorem~\ref{thm:RIP} \emph{also holds} for the matrix $R\Phi^{\dagger}\Psi$ even though only $R\Phi^{\dagger}$ was chosen randomly. In this sense, the random matrix models above form ``universal encoders" because, if the RIP property holds for $R\Phi^{\dagger}$, then it also holds for $R\Phi^{\dagger}\Psi$ \emph{independently} of $\Psi$. 
So, as long as the signal is sparse in some basis that is known a priori then, with high probability, we can recover the signal using a random matrix model and convex optimization techniques. Second, this theorem is equivalent to Theorem 5.2 of Ref.~\cite{Baraniuk2007}, where they fixed $m$, $n$, and $\delta$ and deduced upper bounds on the sparsity (therefore the constant $C_1$ above is the inverse of $C_1$ in Ref.~\cite{Baraniuk2007}). 

Some common examples of probability distributions that lead to the concentration inequality in Eq.~(\ref{eq:concentration}) are
\begin{enumerate}
\item Sampling the $n$ columns uniformly at random from $\mathbb{S}^{m-1}$,
\item Sampling each entry from a normal distribution with mean 0 and variance $\frac{1}{m}$,
\item Sampling the $m$ rows by random $m$-dimensional projections $P$ in $\mathbb{R}^n$ (and normalizing by $\sqrt{\frac{n}{m}}$),
\item Sampling each entry from the symmetric Bernoulli distribution $\mathbbm{P}\left(A_{i,j}=\pm\frac{1}{\sqrt{m}}\right) = \frac{1}{2}$,
\item Sampling each entry from from the set $\left\{-\sqrt{\frac{3}{m}},0,\sqrt{\frac{3}{m}}\right\}$ according to the probability distribution $\left\{\frac{1}{6},\frac{2}{3},\frac{1}{6}\right\}$.
\end{enumerate}
Example 4 will be of particular importance for our main results. We now analyze how to relax the condition of sparsity to compressibility and also how to take noise effects into account.

\subsection{Compressive Sensing in the Presence of Noise}\label{sec:CS3}

The last part of this brief outline of CS techniques shows that they are robust to both small deviations from sparsity as well as noise in the signal. First, for any vector $x$, let $x_S$ represent the vector resulting from only keeping the $S$ entries of largest magnitude. Thus, $x_S$ is the best $S$-sparse approximation to $x$ and $x$ is called ``$S$-compressible" if $\|x-x_S\|_1$ is small. In many applications, the signal of interest $x$ only satisfies the property of compressibility, since many of its coefficients can be small in magnitude but are not exactly equal to 0. In addition, real signals typically are prone to noise effects. Suppose the signal and measurement processes we are interested in are affected by noise that introduces itself as an error $\epsilon$ in the measurement vector $y$. We have the following convex optimization problem which includes the noise term of strength $\epsilon$:

\medskip
\noindent
\underline{Convex Optimization Problem (COP~3)} \vspace{4pt}\\
$\textrm{argmin}  \{\|\tilde{x}\|_1 \: : \: \tilde{x} \in \mathbb{R}^n\} \: \: \text{subject to}\:\:  \|y-A\tilde{x}\|_2 \leq \epsilon$.

\medskip

\noindent We now have the following CS theorem~\cite{Foucart2010}, which is the most general form that we will be considering.

\begin{theorem}\label{thm:CompressiveRIP3}

If the matrix $A$ satisfies the RIP of order $2S$, and $\delta_{2S} <  0.4652$, then the solution $x^*$ of the COP~3 satisfies
\begin{align}
\|x^*-x\|_2 &\leq \frac{C_3\|x-x_S\|_1}{\sqrt{S}} +\epsilon,
\end{align}
where $x_S$ is the $S$-compressed version of $x$.
\end{theorem}

\noindent Theorem~\ref{thm:CompressiveRIP3} is deterministic, holds for \emph{any} $x$, and says that the recovery is robust to noise and is just as good as if one were to only keep the $S$ largest entries of $x$ (the compressed version of $x$). If the signal is exactly $S$-sparse then $x=x_S$ and the recovery is exact, up to the noise term $\epsilon$. We now discuss how to apply CS theory to dynamic parameter estimation and quantum magnetometry.

\section{Quantum Dynamic Parameter Estimation with Compressive Sensing}\label{sec:Main1}

We now present the main results of the paper, combining ideas from CS presented above with coherent control of quantum sensors. For concreteness, we adopt notation suitable for quantum magnetometry. Thus, we assume the deterministic function of interest is a magnetic field $b(t)$ that we want to reconstruct on the time interval $[0,T]$. We partition $I=[0,T]$ into $n$ uniformly spaced intervals with endpoints $t_j=\frac{jT}{n}$ for $j \in \{0,...,n\}$ and discretize $[0,T]$ into the $n$ mid-points $\{s_j\}_{j=0}^{n-1}$ of these intervals,
\begin{align}
s_j&=\frac{(2j+1)T}{2n}.
\end{align}
The discretization of $b(t)$ to a vector $B \in \mathbb{R}^n$ is defined by $B_j = b(s_j)$ for each $j \in \{0,...,n-1\}$.  In principle, each $B_j$ can be approximately estimated by performing a Ramsey protocol in each time interval and assuming the magnetic field is constant over the partition width $\delta=\frac{T}{n}$. The result of the Ramsey protocol is to acquire $\frac{1}{\delta}\int_{t_j}^{t_{j+1}}b(t)dt$. The key idea is that, instead of this naive method that requires $n$ measurements, we can apply control sequences during the total time $T$ to modulate the evolution of $b(t)$ before making a measurement at $T$. While we still need to repeat the measurement for different control sequences, this method is amenable to CS techniques, with the potential for exponential savings in the number of measurements needed for an accurate reconstruction of $B$.

Using coherent control to reconstruct simple sinusoidal fields was performed in Ref.~\cite{TCC} by using the Hahn spin-echo~\cite{Hahn50} and its extensions, such as the periodic dynamical decoupling (PDD) and Carr-Purcell-Meiboom-Gill (CPMG) sequences~\cite{CP,MG}. Recently, control sequences based on the Walsh basis~\cite{Walsh23} have been proposed to reconstruct fields of a completely general form in an accurate manner~\cite{CMYC}. The main point in all of these methods is to control the system during the time $[0,T]$ by performing $\pi$ rotations at pre-determined times $t_j \in [0,T]$. Let us briefly describe the control and measurement processes.

At each $t_j$, a $\pi$-pulse is applied according to some pre-defined algorithm encoded as a length-$n$ bit string $u$. The occurrence of a $1$ ($0$) in $u$ indicates that a $\pi$-pulse should (should not) be applied. The evolution of the system is then given by
\renewcommand \arraystretch{2.5}
\begin{equation}\begin{array}{ll}
U(T)&= \displaystyle\prod_{j=n-1}^{0}\left[\left(e^{-i\int_{t_j}^{t_{j+1}}b(t)dt}\sigma_z\right)\pi^{u(j)}\right]  \\
&= e^{-i\left[\int_0^T\kappa_u(t) b(t)dt\right] \sigma_z}= e^{-iT\langle \kappa_u(t),b(t)\rangle \sigma_z},
\end{array}\label{eq:kappaU}
\end{equation}
where $\kappa_u(t)$ is the piecewise constant function taking values $\pm 1$ on each $(t_j,t_{j+1})$ and a switch $1 \leftrightarrow -1$ occurs at $t_j$ if and only if a $\pi$-pulse is implemented at $t_j$ (we assume without loss of generality that $\kappa_u(t) = 1$ for $t<0$). 
The value of $\kappa_u$ on an interval $[t_j,t_{j+1})$, $j\in\{0,...,n-1\}$, is determined by the parity of the rank of the truncated sequence $u^j=(u(0),...,u(j))$,
\begin{align}
\kappa_u\left[ (t_j,t_{j+1}) \right] = (-1)^{\text{parity}\left(\sum_{i=0}^ju(t_i)\right)}.
\label{eq:kappa}
\end{align}
Hence, performing a $\pi$-modulated experiment produces a phase evolution given by
\begin{align}
\phi_u(T)&=T\langle \kappa_u(t),b(t)\rangle = \int_0^T\kappa_u(t) b(t)dt.
\label{eq:phi}
\end{align}
Performing a measurement in the computational basis $\{|0\rangle, |1\rangle\}$ gives the following probability of obtaining outcome ``0"
\begin{align}
p_0&=\frac{1+\sin\left(T\langle \kappa_u(t),b(t)\rangle\right)}{2}.
\end{align}
Hence, for each $u$, one can solve for $\langle \kappa_u(t),b(t)\rangle$ by estimating $p_0$ and solving for $\langle \kappa_u(t),b(t)\rangle$.

If the set of all $\kappa_u$ form an orthonormal basis for the set of square-integrable functions on $[0,T]$, denoted $L^2[0,T]$, then we can write
\begin{equation}
b(t)=\sum_u \langle \kappa_u(t),b(t)\rangle \kappa_u(t).
\end{equation}
We know that the $\kappa_u$ are piecewise continuous functions and take the values $\pm 1$. An example of a piecewise continuous orthonormal basis of $L^2[0,T]$ is the Walsh basis~\cite{Walsh23}, which we denote $\{w_m\}_{m=0}^\infty$ (see Appendix \ref{sec:Walsh} for details). Each $m$ corresponds to a different control sequence and one can reconstruct $b(t)$ according to its Walsh decomposition
\begin{equation}
b(t)=\sum_{m=0}^\infty \langle w_m(t),b(t)\rangle w_m(t).
\end{equation}
The Walsh basis will be useful for our first set of results in Sec.~\ref{sec:Main1} where we use incoherent measurement bases to reconstruct $b(t)$. 

On the other hand, the set of all $\kappa_u$ need not be a basis. They can be chosen to be random functions, which are also useful from the point of view of random matrices and the RIP. We use randomness and the RIP for our second set of results in Sec.~\ref{sec:Main2}, where $b(t)$ can be sparse in \emph{any} basis.

\subsection{Reconstructing Temporally Sparse Magnetic Fields Using Incoherent Measurement Bases}\label{sec:Main1b}

We first focus  on sparse time domain signals, which are important since they can model real physical phenomena such as action potential pulse trains in neural networks~\cite{DA}. In this case, the parameter $b(t)$ has an $S$-sparse discretization $B$ when written in the standard spike basis of $\mathbb{R}^n$. The spike basis is just the standard basis that consists of vectors with a ``1" in exactly one entry and ``0" elsewhere. To keep notation consistent, we denote the spike basis by $\Psi$. From Theorem~\ref{thm:Compressive1}, if we want to reconstruct $B$ using a measurement basis $\Phi$, then we need  $\Psi$ and $\Phi$ to be incoherent. It is straightforward to show that the discrete orthonormal Walsh basis $\{W_j\}_{j=0}^\infty$ (see Appendix~\ref{sec:Walsh}) is maximally incoherent with the spike basis. Thus, let us suppose that the measurement basis $\Phi$ is the discrete Walsh basis in sequency ordering~\cite{Walsh23}. The Walsh basis is particularly useful because it can be easily implemented experimentally~\cite{CMYC} and it has the added advantage of refocusing dephasing noise effects~\cite{HKV}.

In order to estimate the $k$'th coefficient $B_k^\Phi$ one needs to apply control sequences that match the $k$'th Walsh function. More precisely, for $j \in \{1,...,n-1\}$, if a switch $+1 \leftrightarrow -1$ occurs in the $k$'th Walsh function at time $t_j$ then a $\pi$-pulse is applied at $t_j$. We further assume that available resources limit us to implementing Walsh sequences of order $N$ so that $n=2^N$. Practically, the resources that determine the largest possible $n$ one could use depends on the situation. We are therefore constrained to the information contained in the discretization of $b(t)$ to the vector $B \in \mathbb{R}^{2^{N}}$. The discretized magnetic field $B$ is given in the $\Psi$ and $\Phi$ bases by
\begin{align}
B&=\displaystyle \sum_{k=0}^{2^{N-1}} B_k^{\Psi}\psi_k \!=\! \sum_{k=0}^{2^{N-1}} b(s_k)\psi_k \!=\! \sum_{k=0}^{2^{N-1}} B_k^{\Phi}\Phi_k  \\
&=\! \displaystyle\sum_{k=0}^{2^{N-1}} \langle B,\Phi_k\rangle \Phi_k\!=\! \sum_{k=0}^{2^{N-1}}\! \left[   \sum_{j=0}^{n-1}b(s_j)\frac{w_k(s_j)}{\sqrt{n}}\right]\!\Phi_k.\nonumber
\end{align}

From Theorem~\ref{thm:Compressive1} we expect that, since $B$ is assumed to be sparse in the standard basis, very few measurements of Walsh coefficients (much smaller than $2^N$) are sufficient to reconstruct $b$ with high probability. Let us rephrase Theorem~\ref{thm:Compressive1} in the notation introduced here.

\begin{theorem}\label{thm:Compressive2}

Let $B=(b(s_0),...,b(s_{n-1})) \in \mathbb{R}^{n}$ be the $N$'th order discretization $b(t) \in [0,T]$ ($n=2^N$) and suppose $B$ is $S$-sparse in the spike basis $\Psi$. Let us select $m$ measurements (discrete Walsh vectors) uniformly at random from $\Phi$ and denote the set of chosen indices by $M$ ($|M|=m$). Then, if $\delta > 0$ and
\begin{equation}
m\geq C S \log\left(\frac{n}{\delta}\right),\label{eq:mbound1}
\end{equation}
the solution to the following convex optimization problem is equal to $B^{\Psi}$ with probability no less than $1-\delta$,

\medskip

\noindent
\underline{Convex Optimization Problem} \vspace{4pt}\\$\textrm{argmin}  \{\|x\|_1 \: : \: x \in \mathbb{R}^n\} \: \: \text{subject to} : \: \: \forall j \in M,\: \: $
\begin{align}
B_j^{\Phi}&=\left\langle \phi_j,\sum_{j=0}^{n-1}x_j\psi_j\right\rangle.
\end{align}
\end{theorem}

\medskip

\noindent Note that Theorem~\ref{thm:Compressive2} also holds in the case where there is noise in the measurement results $B_j^{\Phi}$.

An important point that needs to be clarified is that real measurements performed using a spin system such as the NV center produce the coefficients of $b(t)$ with respect to the \emph{continuous} Walsh basis $\{w_k(t)\}_{k=0}^\infty$. These coefficients, which we denote by $\hat{b}_k$, are not equal to the coefficients $B_k^\Phi$. Thus, to be completely rigorous, we need to understand how to compare $\hat{b}_k$ and $B_k^\Phi$. It will be more useful to multiply $\hat{b}_k$ by $\sqrt{n}$ and compare the distance between $\sqrt{n}\hat{b}_k$ and $B_k^\Phi$. If we measure $m$ coefficients using the spin system and obtain a vector $y = (\hat{b}_{\alpha_1},...,\hat{b}_{\alpha_m})$, we can then bound the distance between $y$ and $z=Ax$ where 
\begin{equation}
z_{\alpha_k}=B_{\alpha_k}^\Phi.
\end{equation}
Then, if we obtain 
\begin{equation}
\|y-z\|_2\leq \epsilon,
\end{equation}
for some $\epsilon > 0$, we can treat the vector $y$ obtained from the real sensing protocol as a \emph{noisy version} of $z$. Since CS is robust to noise, the reconstruction will still be successful if $\epsilon$ is small. So, let us bound $\|y-z\|_2$ by first bounding the distance between $\sqrt{n}\hat{b}_k$ and $B_k^\Phi$ for some $k$.
We have
\begin{align}
\hat{b}_k &= \frac{1}{T}\int_0^T b(t) w_k(t)dt = \frac{1}{T}\int_0^T g_k(t)dt,
\end{align}
where we let $w_k$ denote the scaled version of the $k$'th Walsh function to $[0,T]$ and we define the function $g_k$ by $g_k(t)=b(t)w_k(t)$. Now
\begin{align}
B_k^\Phi&=\frac{1}{\sqrt{n}}\sum_{j=0}^{n-1}w_k(s_j)b(s_j)=\frac{1}{\sqrt{n}}\sum_{j=0}^{n-1}g_k(s_j),
\end{align}
and so
\begin{align}
\left|\sqrt{n}\hat{b}_k - B_k^\Phi\right| &= \left|\frac{\sqrt{n}}{T}\int_0^T g_k(t)dt - \frac{1}{\sqrt{n}}\sum_{j=0}^{n-1}g_k(s_j)\right|\nonumber \\
&=\frac{\sqrt{n}}{T}\left|\int_0^T g_k(t)dt - \frac{T}{n}\sum_{j=0}^{n-1}g_k(s_j)\right|.\nonumber \\
\end{align}
By the midpoint error formula for approximating integrals by Riemann sums~\cite{Spivak} we have
\begin{align}
\left|\sqrt{n}\hat{b}_k - B_k^\Phi\right| &\leq \frac{\sqrt{n}}{T}\text{max}_{t\in [0,T]}\left|b^{''}(t)\right|\frac{T}{24}\frac{T^2}{n^2}\nonumber \\
&= \frac{T^2}{24n^{\frac{3}{2}}}\text{max}_{t\in [0,T]}\left|b^{''}(t)\right|,
\end{align}
and so, for $y$ and $z$ defined above,
\begin{align}
\|y-z\|_2\leq \frac{\sqrt{m}T^2}{24n^{\frac{3}{2}}}\text{max}_{t\in [0,T]}\left|b^{''}(t)\right|.
\end{align}
Thus, we can set 
\begin{align}
\epsilon&=\frac{\sqrt{m}T^2}{24n^{\frac{3}{2}}}\text{max}_{t\in [0,T]}\left|b^{''}(t)\right|,
\end{align}
which is small since CS requires $m \sim O\left(S\log(n)\right)$ measurements, where $S$ is the sparsity level. Thus, we have
\begin{align}
\epsilon&\sim\frac{T^2}{24}\text{max}_{t\in [0,T]}\left|b^{''}(t)\right| \sqrt{\frac{S\log(n)}{n^3}},
\end{align}
which converges to 0 quickly in $n$. Hence, using the coefficients $\hat{b}_k$ obtained from the physical sensing protocol will still provide robust reconstructions of the magnetic field. We now use these results to discuss applications to neural magnetometry, with the goal being to reconstruct the magnetic field profile of firing action potentials in neurons.



\subsection{Numerical Simulations}\label{sec:Numerics}

Here, we present a numerical analysis of using CS to reconstruct time-sparse magnetic fields. Since our only constraint is that the signal is sparse in the standard basis, there is clearly a wide range of possible models we can choose from. To place the model within the context of a physically relevant scenario, we assume the magnetic field is a series of two-sided spikes as is the case when a series of action potentials is produced by a firing neuron~\cite{WRW,CDF,ACCP,MO}.
There is large variation in the physical parameters describing such neuronal activity. We chose a set of parameters that both aid in highlighting the potential advantages of CS and are close to current specifications for a sensing system such as the NV center~\cite{Pham11, HBT,CMYC}. We assumed a total acquisition time of $T=1$ ms and defined an ``event" to be a single action potential, which we assumed to last 10$\mu$s. As well, we assumed that five events occur in the 1~ms time-frame and that control pulse times last approximately 10~ns. 
We chose these parameters, which  lie at the current extremes of each system in order  to have many different events occurring in [0,T]  (see figures below). Parameters closer to current experimental capabilities (e.g.  pulse times of 40 ns and action potentials of 100$\mu$s) would have resulted in a smaller number of events and thus less meaningful numerical comparisons.

Each of the two spikes in an action potential was assumed to have maximum magnitude of 1 nT and last $\Delta=5~\mu$s. If $\tau_P$ denotes the pulse time then, in practice, one only has to choose a reconstruction order $N$ such that the resolution $\frac{1}{n}=2^{-N}$ of the partition of size $2^N$ on the time interval $[0,T]$ satisfies the following two-sided inequality condition
\begin{align}
\tau_P <  2^{-N} T < \Delta.
\end{align}
Hence, we can take $N=10$ (so $n=2^{10}$) as a suitable reconstruction order. More precisely, $2^{-10}$ms $\sim 1 \: \mu $s is a fine enough resolution to capture events of length $\Delta$, yet coarse enough so that $10$ns pulses can be approximated as $\delta$-pulses. For $n=1024$, the average and maximum number of pulses one has to apply in a sequence are 512 and 1024 respectively. Since CS techniques are robust to noise, we expect the reconstructions to be relatively robust against pulse errors and imperfections.

We implemented a CS reconstruction using random, non-adaptive, measurements in the Walsh basis. We emphasize that the Walsh and spike bases are maximally incoherent and that the Walsh measurement basis is easily implemented in practice. Again, the number of events in the spike train was chosen to be 5 so that there are 10 total magnetic field pulses (5 of each polarity). For simplicity, the times of each active potential event were chosen uniformly at random in $[0,T]$. We therefore have a sparsity level of $S=50$ and so the number of measurements $m$ we should require is
\begin{align}
m &\sim O\left(S\log\left(n\right)\right)\nonumber \\
&\sim O(500).
\end{align}
As mentioned in Sec.~\ref{sec:CS} there is an empirical rule that says in many cases around $4S$ or $5S$ measurements typically suffice to reconstruct the signal. Thus, we might expect around 200 or 250 measurements are adequate to reconstruct the signal with high probability.

The protocol was numerically implemented for various values of $m$ up to $500$. Once $m$ reached 200, as expected by the above empirical rule, the probability of successful exact reconstruction, $\text{p}_{\text{suc}}$, began to quickly converge to 1 which verifies that the empirical rule is satisfied here. At $m=250$, the probability was essentially equal to 1. For each $m \in \{200,210,220,...,290,300\}$ we recorded $\text{p}_{\text{suc}}$ from a sample of 1000 trials. Since the reconstruction is either close to exact or has observable error, and the majority of the actual signal in the time-domain is equal to 0 (which implies the total variation of the signal is small), we set a stringent threshold for determining successful reconstruction; if the mean-squared error (MSQE) between the simulated ($\Bsim$) and reconstructed ($\Brec$) magnetic fields was less than $10^{-9}$ (nT)$^2$s then the reconstruction was counted as successful. The results are contained in Table~\ref{table:1}.

\begin{figure}[t]\begin{center}
\includegraphics[width=.45\textwidth]{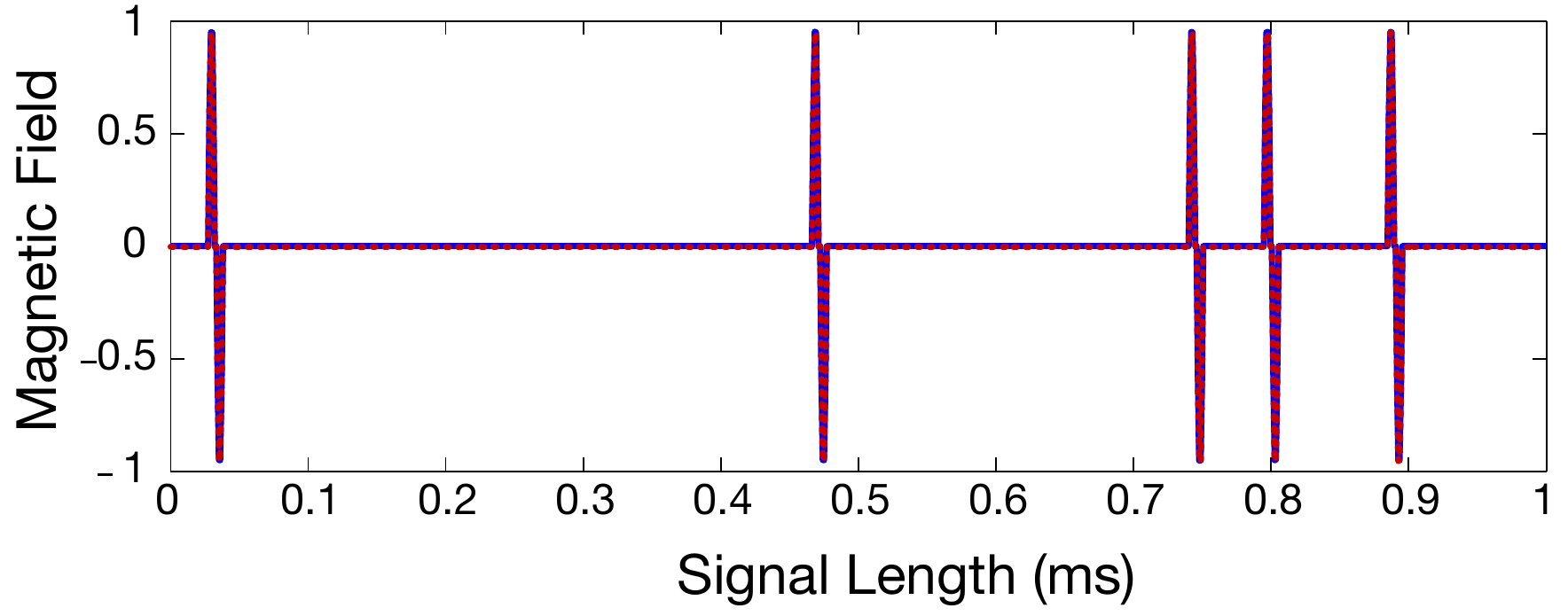}
\caption{\label{Fig:CSsuc} Simulated (blue solid) and Successful CS Reconstructed (red dotted) Magnetic Fields (5 events with m=200 and MSQE=7.6109e-12)}
\end{center}
\end{figure}

\begin{figure}[t] \begin{center}
\includegraphics[width=.45\textwidth]{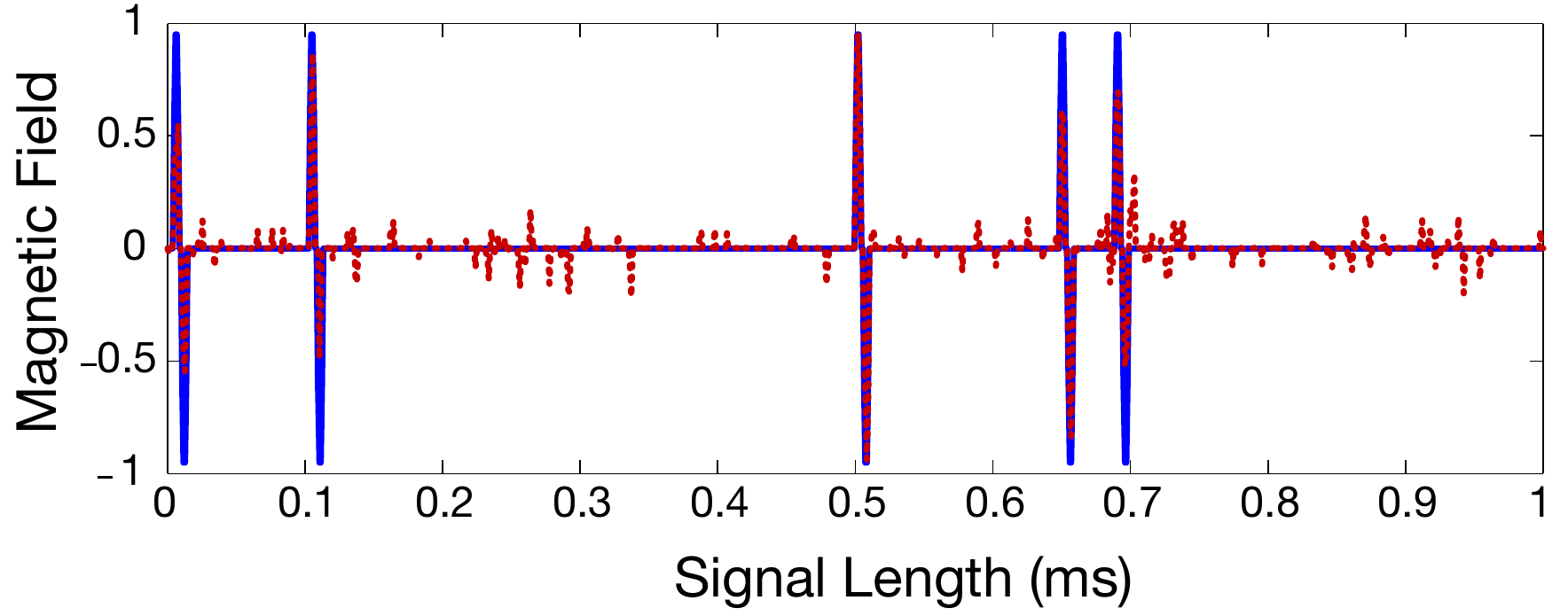}
\caption{\label{Fig:CSunsuc} Simulated (blue solid) and Unsuccessful CS Reconstructed (red dotted) Magnetic Fields (5 events with m=200 and MSQE=0.0051747).}
\end{center}
\end{figure}

The main message is that, if the relevant experimental parameters are taken into account, one obtains a range of possible values for $n$ which defines the required resolution of the grid on $[0,T]$. This provides an upper bound on the number of operations in a Walsh sequence that have to be performed. If $S$ is the expected sparsity of the signal with respect to the chosen value for $n$ then taking
$m \sim O\left(S\log\left(n\right)\right)$
measurements implies the probability of successful reconstruction, $\text{p}_{\text{suc}}$, will be high and converges to 1 quickly as $m$ grows larger. We plotted examples of successful and unsuccessful reconstructions in Fig.~\ref{Fig:CSsuc} and~\ref{Fig:CSunsuc} respectively. Typically, the CS reconstruction either works well (MSQE $< 10^{-9}$) or clearly ``fails" (MSQE $ \sim 0.01$).

\begin{table*}
\centering
    \begin{tabular}{| l | l | l | l | l | l | l | l | l | l | l | l |}
    \hline
    m & 200 & 210 & 220 & 230 & 240 & 250 & 260 & 270 & 280 & 290 & 300 \\ \hline
   $\text{p}_{\text{suc}}$ & 0.870 & 0.950 & 0.974 & 0.991 & 0.996 & 0.999 & 1.000 & 1.000 & 1.000 & 1.000 & 1.000\\
    \hline
    \end{tabular}
     \caption{\label{table:1} Probability of successful CS reconstruction, $\text{p}_{\text{suc}}$, for different values of $m \ll n=1024$.}
\end{table*}

\subsubsection{Accounting for Decoherence}\label{sec:Noise}

From Theorem~\ref{thm:CompressiveRIP3}, we know that CS techniques are robust in the presence of noise. We model the noise and evolution of the system as follows. We suppose that the Hamiltonian of the system is given by
\begin{align}
\hat{H}(t)&=\left[b(t)+\beta(t)\right]\sigma_z,
\end{align}
where $\beta(t)$ is a zero-mean stationary stochastic process. 
By the Wiener-Khintchine theorem, the spectral density function of $\beta(t)$, $S_\beta(\omega)$, is equal to the Fourier transform of the auto-correlation function of $\beta(t)$ and thus the decay in coherence of the quantum system is given by $v=e^{-\chi(t)}$, where 
\begin{align}
\chi(T)&=\int_0^\infty\frac{S_\beta(\omega)}{\omega^2}F(\omega T),\label{eq:chidef}
\end{align}
and $F(\omega T)$ is the filter function for the process~\cite{Uhrig08}. It is important to note that one can experimentally reconstruct $S_\beta(\omega)$ by applying pulse sequences of varying length and inter-pulse spacings that match particular frequencies~\cite{CLL,BGY,BPB}.

Applying control sequences that consist of $\pi$-pulses during $[0,T]$ modulates $\chi(T)$ by modifying the form of $F(\omega T)$ in Eq.~(\ref{eq:chidef})~\cite{Uhrig08}. In most experimentally relevant scenarios, low frequency noise gives the most significant contribution to $S_\beta(\omega)$. Hence, typically, the goal is to design control sequences that are good high-pass filters. When the control sequence is derived from the $j$'th Walsh function, we have
\begin{align}
\chi_j(T)&=\int_0^\infty\frac{S_\beta(\omega)}{\omega^2}F_j(\omega T),
\end{align}
where $F_j(\omega T)$ is the filter function associated with the $j$'th Walsh function.  The low frequency behavior of each $F_j$ has been analyzed in detail~\cite{HKV,MCYC}; in general, if the noise spectrum is known,  one can obtain an estimate of each $\chi_j(T)$, and thus each $v_j$.

The signal $S_j$ acquired from a sensing protocol with the $j$'th Walsh control sequence is given by
\begin{align}
S_j(T)&=\frac12\left[1+v_j(T)\sin(z_j(T))\right],
\end{align}
where
\begin{align}
z_j(T)&=\frac{1}{T}\int_0^Tw_j(t)b(t)dt.
\end{align}
We note that for zero-mean noise we have
\begin{align}
\left\langle\int_0^Tw_j(t)\beta(t)dt\right\rangle&=0.
\end{align}
Now, for $N_j$ measurement repetitions, we have that the sensitivity in estimating $z_j$, denoted $\Delta z_j$, is given by~\cite{MCYC}
\begin{align}
\Delta z_j &=\frac{1}{\sqrt{N_j}Tv_j}.
\end{align}
Thus, fluctuations in obtaining the measurement results $z_j$ are on the order of $\frac{1}{\sqrt{N_j}Tv_j}$ and so the $2$-norm distance between the ideal measurement outcome vector $z$ and actual measurement vector $y$ is
\begin{align}
\|\vec{z}-\vec{y}\|_2\sim \sqrt{\sum_{j=1}^m \left[\frac{1}{\sqrt{N_j}Tv_j}\right]^2} = \sqrt{\sum_{j=1}^m \frac{1}{N_jT^2v_j^2}}.
\end{align}
Since CS reconstructions are robust up to $\|\vec{z}-\vec{y}\|_2$, we have that a realistic CS reconstruction in a system such as the NV center is robust up to $\sqrt{\sum_{j=1}^m \frac{1}{N_jT^2v_j^2}}$.

 \section{Reconstructing General Sparse Magnetic Fields} \label{sec:Main2}

While time-sparse signals are important in various applications, they are far from being common. Ideally, we would like to broaden the class of signals that we can reconstruct with CS techniques and quantum sensors. In this section, we present a method for reconstructing signals that are sparse in \emph{any} known basis. This generality of the signal is significant since most real signals can be approximated as sparse in some basis. We use the results of Sec.~\ref{sec:CS2} to show that performing \emph{random} control sequences creates random measurement matrices that satisfy the RIP (see Def.~\ref{def:RIP}) with high probability. Therefore, by Theorem~\ref{thm:CompressiveRIP1}, a small number of measurements suffice for exact reconstruction of the signal. Again, we emphasize the key point that the signal can be sparse in \emph{any} basis of $\mathbb{R}^n$.

\subsection{Applying Random Control Sequences}\label{sec:Random}

Suppose $b(t)$ is a deterministic magnetic field on $I=[0,T]$ and we partition $I$ into $n$ uniformly spaced intervals with grid-points $t_j=\frac{jT}{n}$ for $j \in \{0,...,n\}$. Previously, we used the discrete Walsh basis $\{w_k\}_{k=0}^{2^N-1}$ as our measurement basis $\Phi$. $\pi$-pulses were applied at each $t_j$ according to whether a switch between $+1$ and $-1$ occurred at $t_j$. Now, for each $j\in \{0,...,n-1\}$, we choose to either apply or not apply a $\pi$-pulse at $t_j$ according to the symmetric Bernoulli distribution
\begin{equation}
\mathbbm{P}\left[\pi \: \: \text{applied at} \:\: t_j\right]=\frac{1}{2}.
\end{equation}
The result of this sampling defines  the $n$-bit string $u$, where the occurrence of a $0$ indicates a $\pi$-pulse is not applied, and a $1$ indicates a $\pi$-pulse is applied. Following Eq.~(\ref{eq:kappaU})-(\ref{eq:phi}), the evolution of the system is 
\begin{align}
U(T)&=
e^{-iT\langle \kappa_u(t),b(t)\rangle \sigma_z}.
\end{align}
As before, let us discretize $[0,T]$ into $n$ points $\{s_j\}_{j=0}^{n-1}$ where
\begin{align}
s_j&=\frac{(2j+1)T}{2n}.
\end{align}
Define $\tilde{\kappa}_u$ to be the random element of $\{-1,1\}^n$ with entries given by $\kappa_u(s_j)$ for each $j\in\{0,...,n-1\}$. As well, define $B \in \mathbb{R}^n$ to be the discretization of $b(t)$ at each $s_j$, $B_j=b(s_j)$.

Suppose there is an orthonormal basis $\Psi$ of $\mathbb{R}^n$ for which $B$ has an approximate $S$-sparse representation.
 To maintain notational consistency, we use $\Psi$ to represent the $n\times n$ matrix whose columns are the basis elements $\psi_j \in \mathbb{R}^n$. Let $x$ denote the coordinate representation of $B$ in $\Psi$,
\begin{align}
B= \sum_{j=1}^{n}x_j\psi_j.
\end{align}
We emphasize that $\Psi$ is an \emph{arbitrary} but known basis. Assume we have chosen $m$ random measurement vectors $\kappa_{u_i}$, $i=1,...,m$. and let us define the matrix $\mathcal{G}$ whose entries are given by
\begin{align}
\mathcal{G}_{i,j}&=\frac{\kappa_{u_i}(s_j)}{\sqrt{m}}=\frac{\tilde{\kappa}_{u_i}(j)}{\sqrt{m}},
\end{align}
where $i\in\{1,...,m\}$ and $j\in \{1,...,n\}$. 
Also, let us define the $m\times n$ matrix
 \begin{align}
 A=\mathcal{G}\Psi.
 \end{align}

Since the entries of $\mathcal{G}$ were chosen according to a probability distribution that satisfies Eq.~(\ref{eq:concentration}), from Theorem~\ref{thm:RIP}, 
 there exist constants $C_1$, $C_2 > 0$ which depend only on $\delta$ such that, with probability no less than 
\begin{equation}
1-2e^{-C_2n},
\end{equation}
$\mathcal{G}$, and thus $A$, satisfies the RIP of order $(2S,\delta)$ when
\begin{equation}
m\geq 2C_1S\log\left(\frac{n}{2S}\right).
\end{equation}
By Theorem~\ref{thm:CompressiveRIP3} this implies that, if we let $y$ be the noisy version of the ideal measurement $Ax$ and $\|y- Ax\|_2 \leq \epsilon$, the solution $x^*$ of COP~3 satisfies,
\begin{align}
\|x^*-x\|_2 &\leq \frac{C_3\|x-x_S\|_1}{\sqrt{S}}+\epsilon
\end{align}
where $x_S$ is the best $S$-sparse approximation to $x$ and $C_3$ is a constant. 

We note that the real sensing procedure in a spin system calculates the continuous integrals of the form $\int_0^T\kappa_{u_i}(t)b(t)dt$ rather than the discrete inner products $\langle \frac{1}{\sqrt{m}} \tilde{\kappa}_{u_i},B\rangle$. To take this into account, let us bound the distance between $\frac{n}{T\sqrt{m}}\int_0^T\kappa_u(t)b(t)dt$ and $\frac{1}{\sqrt{m}}\sum_{j=0}^{n-1}\kappa_u(s_j)b(s_j)$. For ease of notation, let $g_u:[0,T]\rightarrow \mathbb{R}$ be defined by $g_u(t)=\kappa_u(t)b(t)$. We have by the the midpoint Riemann sum approximation~\cite{Spivak},
\begin{align}
\left|\frac{n}{T\sqrt{m}}\int_0^Tg_u(t)dt-\frac{1}{\sqrt{m}}\sum_{j=0}^{n-1}g_u(s_j)\right| \nonumber\\
= \frac{n}{T\sqrt{m}}\left|\int_0^Tg_u(t)dt-\frac{T}{n}\sum_{j=0}^{n-1}g_u(s_j)\right| \\
\leq \frac{n}{T\sqrt{m}} \text{max}_{t\in[0,T]}\left|b^{\prime\prime}(t)\right|\frac{T}{24}\frac{T^2}{n^2} \nonumber\\
= \frac{T^2}{24n\sqrt{m}} \text{max}_{t\in[0,T]}\left|b^{\prime\prime}(t)\right|.\nonumber
\end{align}
Hence, if we make $m$ measurements and obtain a vector $y=(y_{u_1},...,y_{u_m})$ where
\begin{align}
y_{u_j}&=\frac{n}{T\sqrt{m}}\int_0^T\kappa_u(t)b(t)dt,
\end{align}
then
\begin{align}
\|y-Ax\|_2&\leq \frac{T^2\sqrt{m}}{24n} \text{max}_{t\in[0,T]}\left|b^{\prime\prime}(t)\right|.
\end{align}
Setting
\begin{align}
\epsilon &= \frac{T^2\sqrt{m}}{24n} \text{max}_{t\in[0,T]}\left|b^{\prime\prime}(t)\right|,
\end{align}
implies we can treat $y$ as a noisy version of $Ax$,
\begin{align}
\|y-Ax\|_2\leq \epsilon.
\end{align}
Therefore, by Theorem~\ref{thm:CompressiveRIP3} and the fact that $m \sim O\left(S\log\left(\frac{n}{S}\right)\right)$, we have that the CS reconstruction will still be successful up to $\epsilon$ (and becomes exact as $n\rightarrow \infty$). Hence, we can treat the difference between the actual continuous integrals we obtain in $\tilde{y}$ and the ideal discrete measurements $y$ as a fictitious error term in the acquisition process and use Theorem~\ref{thm:CompressiveRIP3}. Finally, if there is actual error $\epsilon_1$ in the magnetic field then taking $\epsilon=\epsilon_1+\tilde{\epsilon}$ and using Theorem~\ref{thm:CompressiveRIP3} again gives that the solution $x^*$ of COP~2 satisfies,
\begin{align}
\|x^*-x\|_2 &\leq \frac{C_3\|x-x_S\|_1}{\sqrt{S}} +\epsilon.
\end{align}

We reiterate that $\Psi$ is arbitrary but known \emph{a priori}. More precisely, in order to implement the convex optimization procedures, one needs to know the basis $\Psi$. However there are no restrictions on what this basis \emph{could} be. If the signal $B$ has an approximately $S$-sparse representation in the basis $\Psi$, then the above result implies we can recover $B$ using a small number of samples.

It is also important to note for the case of when $\Psi$ is the Fourier basis, this result is distinct, and in many cases stronger, than Nyquist's theorem. Nyquist's theorem states that if $B$ has finite bandwidth with upper limit $f_B$, one can reconstruct $\tilde{B}$ by sampling at a rate of $2f_B$. Compressive sensing tells us that even when there is no finite bandwidth of the signal, exact reconstruction is possible using a small number of measurements that depends only on the sparsity level. 

Before analyzing numerical examples, we give a theoretical analysis of how to account for decoherence effects. The general idea is similar  to the discussion in Sec.~\ref{sec:Noise}. The main point is that applying a random control sequence according to the random binary string $u$ gives
\begin{align}
\chi_u(T)&=\int_0^\infty\frac{S_\beta(\omega)}{\omega^2}F_u(\omega T),
\end{align}
where $F_u(\omega T)$ is the filter function associated to this control sequence. In principle, both the nose spectrum $S_\beta(\omega)$ and low frequency behavior of each $F_u(\omega T)$ can be determined~\cite{Uhrig08} so one can estimate $\chi_u(T)$.

The signal $S_u$ one measures from the binary string $u$ is
\begin{align}
S_u&=\frac12\left[v_u(T)\sin(z_u(T))\right],
\end{align}
where
\begin{align}
z_u(T)&=\frac{1}{T}\int_0^T\kappa_u(t)b(t)dt.
\end{align}
Again, noting that an ensemble of measurements is required to acquire the signal above, we have for a zero-average noise $\beta(t)$
\begin{align}
\left\langle\int_0^T\kappa_u(t)\beta(t)dt\right\rangle&=0,
\end{align}
Now, for $N_u$  measurement repetitions we have
\begin{align}
\Delta z_u&=\frac{1}{\sqrt{N_u}Tv_u}.
\end{align}
As before, fluctuations in obtaining the measurement results $z_u$ are on the order of $\frac{1}{\sqrt{N_u}Tv_u}$ and so the $2$-norm distance between the ideal measurement outcome vector $z$ and actual measurement vector $y$ is
\begin{align}
\|\vec{z}-\vec{y}\|_2\sim \sqrt{\sum_u \left[\frac{1}{\sqrt{N_u}Tv_u}\right]^2} = \sqrt{\sum_u \frac{1}{N_uT^2v_u^2}}.
\end{align}
Since CS reconstructions are robust up to $\|z-y\|_2$, we have that a realistic CS reconstruction in a system such as the NV center is robust up to $\sqrt{\sum_u \frac{1}{N_uT^2v_u^2}}$.

\subsubsection{Reconstruction Examples}\label{sec:Numerics1}


We provide a set of numerical examples that show how random measurement matrices allow for reconstructing sparse signals in any known basis $\Psi$. We first analyze the case of $b(t)$ being sparse in the frequency domain and then revisit the case of $b(t)$ being sparse in the time domain (the latter of which can provide a comparison with the results of Sec.~\ref{sec:Numerics}). To keep the discussion both simple and consistent with the example in Sec.~\ref{sec:Numerics}, we choose a total acquisition time of $T=1$ ms, assume pulse widths of $P = 10$ ns, and discretize $[0,T]$ into $n=2^{10}$ intervals.

\medskip

\underline{Frequency-Sparse Signals}


\medskip

Since we have $n=2^{10}$ intervals in $[0,1]$ our high-frequency cut-off is equal to 1.024 MHz. More precisely, a partition of $[0,1]$ into $\frac{1}{n}$ subintervals implies a resolution of $n$ kHz.
We choose a sparsity level $S=8$ with frequencies (in MHz);
\begin{equation}
f_j\in \left\{1.024x_j\right\}_{j=1}^S,
\end{equation}
where the $x_j$ are chosen uniformly at random
\begin{equation}
x_j \in_r \left\{\frac{1}{k}\right\}_{k=1}^{1024}.
\end{equation}
For each randomly chosen frequency, we also choose uniformly random phases $\phi_j \in_r [0,2\pi]$ and amplitudes $A_j \in_r [0,1]$. The resulting field $b(t)$ is given by
\begin{equation}
b(t)=\sum_{j=1}^S A_j\cos\left(2\pi f_j t +\phi_j\right).
\end{equation}
The threshold for the MSQE in this case naturally needs to be larger than for the neural magnetic field reconstruction in Sec.~\ref{sec:Numerics}. The reason for this is that the majority of the signal $b$ in the neuron case is equal to 0 and thus the total variation of that signal is very small (which leads to an extremely small MSQE when successful CS reconstruction occurs). In the case considered here, choosing random frequencies leads to signals with a larger total variation. 

\begin{figure}[thb]\begin{center}
\includegraphics[width=.4\textwidth]{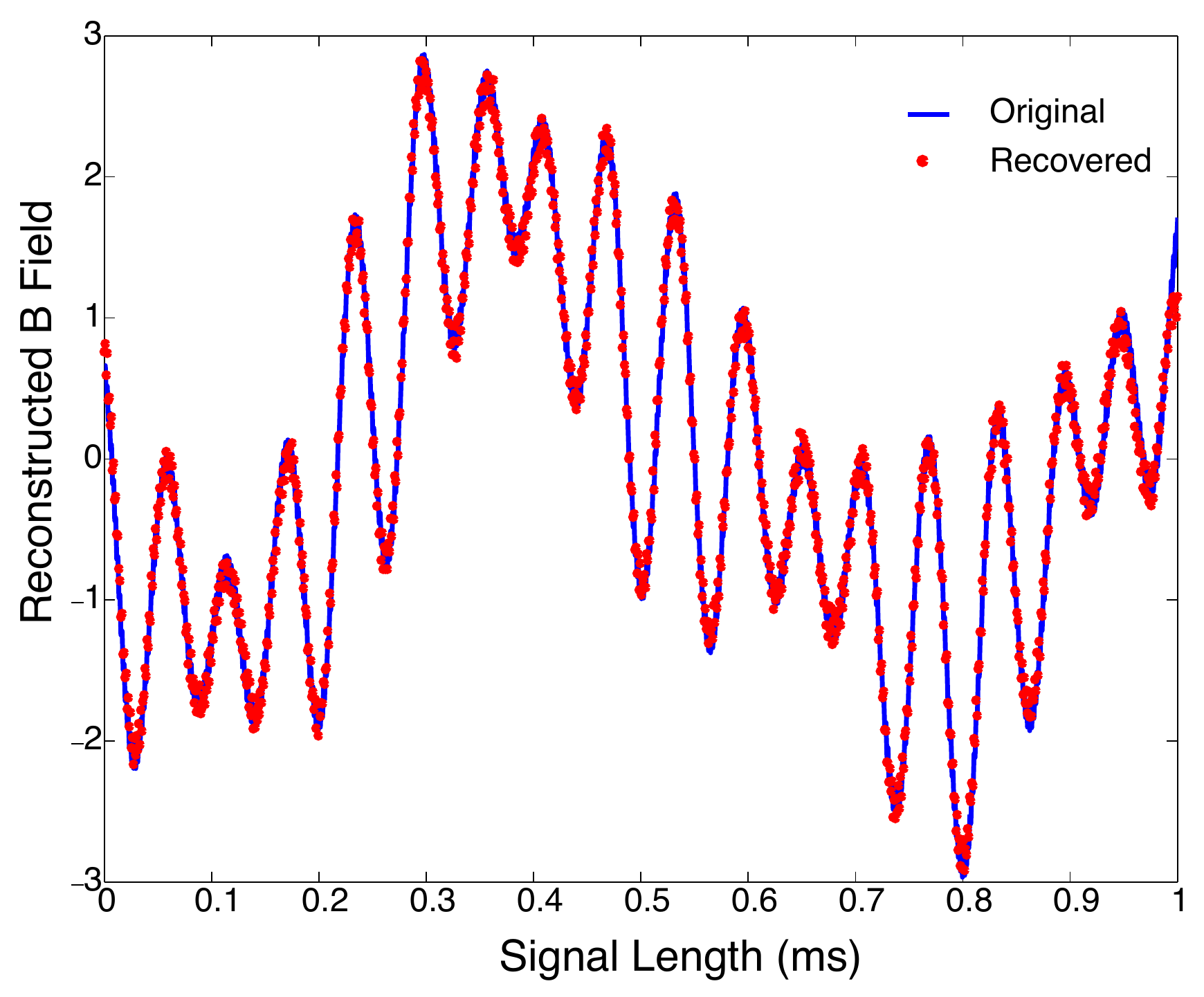}
\caption{\label{Fig:UniversalCSsucEx2rand} CS Reconstruction of (Frequency) Sparse Magnetic Field With 8 Random Frequencies (reconstruction with $m=250$ and MSQE=0.0048006).}
\end{center}
\end{figure}

Fig.~\ref{Fig:UniversalCSsucEx2rand} shows reconstruction for randomly chosen $x_j$ of $\left\{\frac{1}{2},\frac{1}{61},\frac{1}{78},\frac{1}{328},\frac{1}{551},\frac{1}{788},\frac{1}{881},\frac{1}{1022}\right\}$.
\noindent Clearly the reconstruction is fairly accurate. We analyzed the probability of successful CS reconstruction using a MSQE threshold of 0.005 for $m\in\{250,260,...,350\}$. The results are contained in Table~\ref{table:3}. As expected, the probability of successful reconstruction quickly converges to 1 for $m\ll 1024$.
\begin{table*}
\begin{center}
    \begin{tabular}{| l | l | l | l | l | l | l | l | l | l | l | l |}
    \hline
    m & 250 & 260 & 270 & 280 & 290 & 300 & 310 & 320 & 330 & 340 & 350\\ \hline
   $\text{p}_{\text{suc}}$ & 0.914 & 0.958 & 0.983 & 0.993 & 0.998 & 0.999 & 0.999 & 1.000 & 1.000 & 1.000 & 1.000\\
    \hline
    \end{tabular}
    \end{center}
 \caption{\label{table:3} Probability of successful CS reconstruction, $\text{p}_{\text{suc}}$, for different values of $m \ll n=1024$.}
\end{table*}

\medskip

\underline{Time-Sparse Signals}

\medskip

\begin{figure}[thb]
\begin{center}
\includegraphics[width=.45\textwidth]{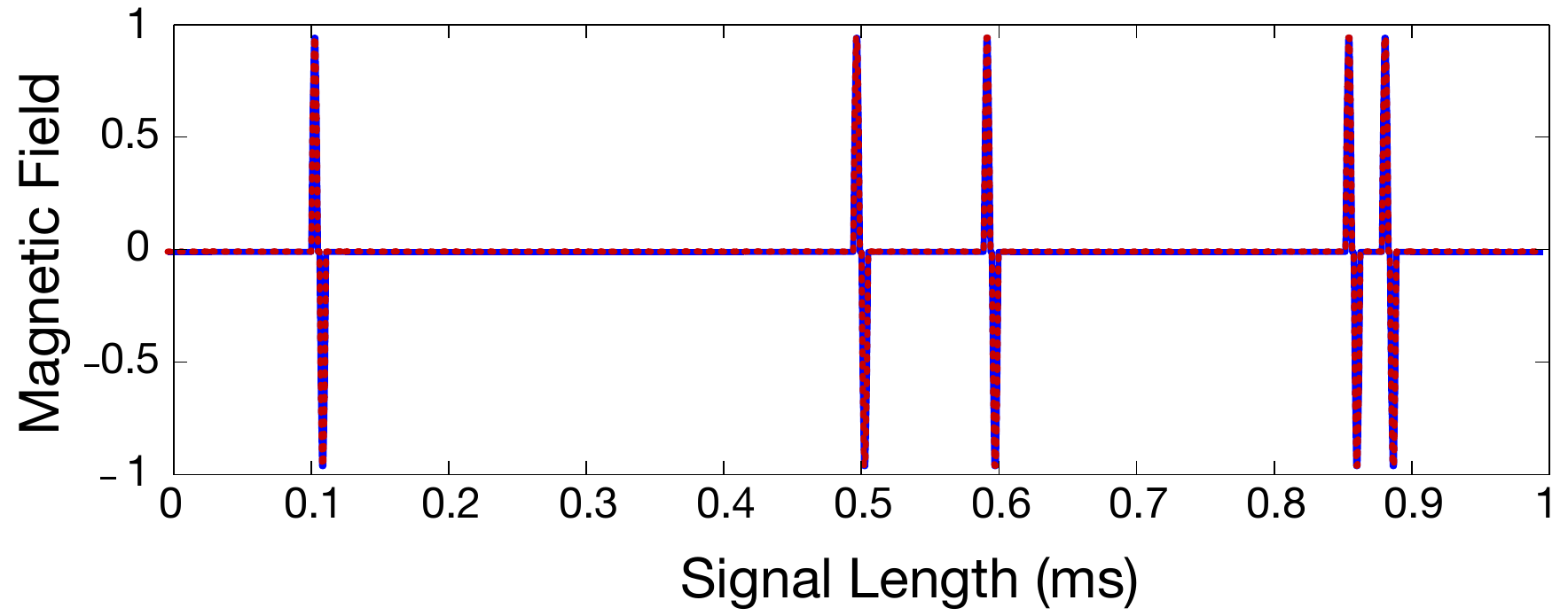}
\caption{\label{Fig:UniversalCSsucneuron} CS Reconstruction of Time Sparse Magnetic Field Produced By Firing Neuron: Simulated (blue solid) and  CS Reconstructed (red dotted) Magnetic Fields (5 events with m=250 random Walsh measurements and MSQE=3.2184$\times10^{-4}$)}
\end{center}
\end{figure}

For this example we used the same parameters as the example in Sec.~\ref{sec:Numerics} however the measurement matrix is now a random Bernoulli matrix of size $m \times n$. Fig~\ref{Fig:UniversalCSsucneuron} shows the original and reconstructed signal for $m=250$ where a MSQE of $\sim 10^{-18}$ was obtained. The probability of successful reconstruction followed a similar form to that in Sec.~\ref{sec:Numerics} in that the probability converged to 1 very quickly as $m$ became larger than 200.

\section{Conclusion}\label{sec:Conclusion}

We have shown that performing quantum magnetometry using the Zeeman effect in spin systems allows one to utilize the power of compressive sensing (CS) theory. The number of measurements required to reconstruct many physically relevant examples of magnetic fields can be greatly reduced without the degradation of the reconstructed signal. This can have significant impact when measurement resources are extremely valuable, as is the case when using quantum systems for reconstructing magnetic fields. In addition, CS is robust to noise in the signal which makes it an extremely attractive tool to use in physically relevant scenarios, especially when phenomena such as phase jitter are present. We have provided a basic introduction to CS by first discussing the concept of incoherent measurement bases and then moving to the random measurement basis picture. The main idea behind CS is to reconstruct sparse signals using both a small number of non-adaptive measurements in some basis and efficient reconstruction algorithms. We have used $l_1$-minimization algorithms however this is only one example of many different methods. 

Our first main example utilized the fact that the Walsh and standard measurement basis are maximally incoherent to model the reconstruction of neural spike trains. The signals are sparse in the time-domain and, since we can perform control pulse sequences that represent Walsh sequences, the signal can be represented using a Walsh measurement matrix. We looked at how the probability of successful reconstruction increased as the number of measurements $m$ increased. The probability saturated at 1 when $m \ll n$, where $\log(n)$ is the chosen Walsh reconstruction order. This order will typically be chosen by the physical parameters of the system used to reconstruct the field. The parameters we chose were relevant for a real system such as the Nitrogen-Vacancy (NV) center in diamond. We also verified that the reconstruction is robust to noise in the signal.

The second main section of our results pertains to random measurements, in particular, random symmetric Bernoulli measurement matrices. These measurement matrices satisfy the Restricted Isometry Property (RIP) with high probability when $m \sim O\left(2S\log\left(\frac{n}{2S}\right)\right)$. We first analyzed signals that are sparse in the frequency domain, and showed that efficient reconstruction occurs with probability 1 for $m \ll n$. In addition, we added a Gaussian noise component and showed successful reconstruction still occurs. We again reiterate that in the case of frequency-sparse signals, CS is completely distinct from Nyquist's Theorem and should be viewed as an independent result. This is easily seen by the fact that we were able to sample at a rate much lower than the highest frequency present in the sample and still obtain an exact reconstruction. Next, we revisited the neuron signals and showed that random measurement matrices are just as effective as the Walsh basis at reconstructing the time-sparse signals.

There are various directions of future research. First, it will be interesting to investigate whether it is possible to incorporate the framework of CS into other methods for performing magnetometry. In this work, we were able to map control sequences directly onto the evolution of the phase which provides the relevant measurement coefficients for efficient reconstruction of the waveform. A different application of CS may be needed for other sensing mechanisms with quantum systems. While we numerically analyzed an example inspired by neuronal magnetometry, there is a wide array of physical phenomena that produce magnetic fields which display sparsity, or more generally compressibility, in some basis. A more detailed analysis of using CS in these cases will be worthwhile to consider.

\begin{appendix}

\section{The Walsh basis}\label{sec:Walsh}

Let $f:[0,T]\rightarrow \mathbb{R}$ be a square-integrable function, that is, $f \in L^2[0,T]$. Then, $f$ has a Walsh representation~\cite{Walsh23} given by
\begin{eqnarray}
f(t)&=& \sum_{j=0}^{\infty}\hat{f}_jw_j(t),
\end{eqnarray}
where
\begin{eqnarray}
\hat{f}_j &=& \frac{1}{T}\int_0^Tf(t)w_j\left(\frac{t}{T}\right)dt, \nonumber
\end{eqnarray}
and the $w_j:[0,1]\rightarrow \mathbbm{R}$, $j \in \mathbb{N}$, denote the Walsh functions. The Walsh functions are useful from a digital viewpoint in that they are a binary system (they only take the values $\pm 1$) and also form an orthonormal basis of $L^2[0,1]$. 

One can explicitly define the Walsh functions via the set of Rademacher functions $\{R_k\}_{k=1}^\infty$. For each $k=1,2,...$, the $k$'th Rademacher function $R_k:[0,1]\rightarrow \mathbb{R}$ is defined by
\begin{equation}
R_k(t)=(-1)^{t_k},
\end{equation}
where $t_k$ is the $k$'th digit of the binary expansion of $t \in [0,1]$,
\begin{align}
t&=\sum_{k=1}^{\infty}\frac{t_k}{2^k} = \frac{t_1}{2}+\frac{t_2}{2^2}+\frac{t_3}{2^3}+.... = 0.t_1t_2t_3.....
\end{align}
%
Equivalently, one can define $R_k$ as the $k$'th square-wave function
\begin{equation}
R_k(t)=\text{sgn}\left(\sin(2^k\pi t)\right).
\end{equation}

The Walsh basis is just equal to the group (under multiplication) generated by the set of all Rademacher functions. Different orderings of the Walsh basis can be obtained by considering the ordering in which one multiplies Rademacher functions together. For instance, two particularly common orderings of the Walsh basis are the ``sequency"~\cite{Walsh23} and ``Paley"~\cite{Paley32} orderings. Sequency ordering arises from writing the set of binary sequences according to ``Gray code"~\cite{Gray53},
which corresponds to having each successive sequence differ in exactly one digit from the previous sequence, under the assumption that right-most digits vary fastest. 
If one assigns each Rademacher function to its corresponding digit in the binary expansion (for instance $R_1$ is associated to the right-most digit, $R_2$ is associated to the next digit and so on)
then, when each integer $i$ is written in Gray code, $i=....i_k....i_2i_1$, we have that the $i$'th Walsh function in sequency ordering is
\begin{align}
w_i(t)&=\displaystyle{\Pi_{k=1}^{\infty}}\left[R_k(t)\right]^{i_k}= \displaystyle{\Pi_{k=1}^{\infty}}(-1)^{i_kt_k}.
\end{align}
Paley ordering of the Walsh basis is obtained in the same manner as just described for sequency ordering, the only difference being that binary sequences are ordered according to standard binary code (rather than Gray code).

There are a couple of important points to remember about Walsh functions. 
First, since sequency and Paley orderings differ only in terms of how each integer $i$ is represented in terms of binary sequences, switching between these orderings reduces to switching between Gray and standard binary code ordering. 
Second, it is straightforward to verify that the set of the first $2^k$ sequency-ordered Walsh functions is equal to the first $2^k$ Paley ordered Walsh functions. Thus, rearrangements of orderings differ only within the sets of functions whose size is a power of $2$.


For the remainder of this paper, whenever the Walsh basis is used, we will assume the functions are sequency-ordered. We define the $n$'th partial sum of $f \in L^2[0,T]$, denoted $f_n$, to be the sum of the first $n$ terms of its Walsh representation
\begin{eqnarray}
f_n(t)&=& \sum_{j=0}^{n-1}\hat{f}_j\Wal_j\left(\frac{t}{T}\right).
\end{eqnarray}
The $N$'th order reconstruction of $f$ corresponds to the $2^N$'th partial sum. 


Lastly, we note that for any $n = 2^N$ with $N\geq 0$, one can also define a discrete Walsh basis for $\mathbb{R}^n$. To see this, first note that the first $2^N$ Walsh functions $\{w_j\}_{j=0}^{2^N-1}$ are piecewise constant on the $2^N$ uniform-length subintervals of $[0,1]$. Now, just associate the values of each $w_j$ on these subintervals to a vector in $\mathbb{R}^n$. The resulting set of vectors is an orthogonal basis, and dividing each vector by $\sqrt{n}$ gives the discrete orthonormal Walsh basis, which we denote by $\{W_j\}_{j=0}^\infty$. As an example, let $N=2$ so $n=4$. Then the four unnormalized Walsh vectors are $\{(1,1,1,1), (1,1,-1,-1), (1,-1,-1,1), (1,-1,1,-1)\}$. Normalizing each vector by 2 gives the orthonormal Walsh basis $\{W_j\}_{j=0}^3$ of $\mathbb{R}^4$.

\end{appendix}


\begin{thebibliography}{51}%
\makeatletter
\providecommand \@ifxundefined [1]{%
 \@ifx{#1\undefined}
}%
\providecommand \@ifnum [1]{%
 \ifnum #1\expandafter \@firstoftwo
 \else \expandafter \@secondoftwo
 \fi
}%
\providecommand \@ifx [1]{%
 \ifx #1\expandafter \@firstoftwo
 \else \expandafter \@secondoftwo
 \fi
}%
\providecommand \natexlab [1]{#1}%
\providecommand \enquote  [1]{``#1''}%
\providecommand \bibnamefont  [1]{#1}%
\providecommand \bibfnamefont [1]{#1}%
\providecommand \citenamefont [1]{#1}%
\providecommand \href@noop [0]{\@secondoftwo}%
\providecommand \href [0]{\begingroup \@sanitize@url \@href}%
\providecommand \@href[1]{\@@startlink{#1}\@@href}%
\providecommand \@@href[1]{\endgroup#1\@@endlink}%
\providecommand \@sanitize@url [0]{\catcode `\\12\catcode `\$12\catcode
  `\&12\catcode `\#12\catcode `\^12\catcode `\_12\catcode `\%12\relax}%
\providecommand \@@startlink[1]{}%
\providecommand \@@endlink[0]{}%
\providecommand \url  [0]{\begingroup\@sanitize@url \@url }%
\providecommand \@url [1]{\endgroup\@href {#1}{\urlprefix }}%
\providecommand \urlprefix  [0]{URL }%
\providecommand \Eprint [0]{\href }%
\providecommand \doibase [0]{http://dx.doi.org/}%
\providecommand \selectlanguage [0]{\@gobble}%
\providecommand \bibinfo  [0]{\@secondoftwo}%
\providecommand \bibfield  [0]{\@secondoftwo}%
\providecommand \translation [1]{[#1]}%
\providecommand \BibitemOpen [0]{}%
\providecommand \bibitemStop [0]{}%
\providecommand \bibitemNoStop [0]{.\EOS\space}%
\providecommand \EOS [0]{\spacefactor3000\relax}%
\providecommand \BibitemShut  [1]{\csname bibitem#1\endcsname}%
\let\auto@bib@innerbib\@empty
\bibitem [{\citenamefont {Giovannetti}\ \emph {et~al.}(2004)\citenamefont
  {Giovannetti}, \citenamefont {Lloyd},\ and\ \citenamefont {Maccone}}]{GLM}%
  \BibitemOpen
  \bibfield  {author} {\bibinfo {author} {\bibfnamefont {V.}~\bibnamefont
  {Giovannetti}}, \bibinfo {author} {\bibfnamefont {S.}~\bibnamefont {Lloyd}},
  \ and\ \bibinfo {author} {\bibfnamefont {L.}~\bibnamefont {Maccone}},\ }\href
  {\doibase 10.1126/science.1104149} {\bibfield  {journal} {\bibinfo  {journal}
  {Science}\ }\textbf {\bibinfo {volume} {306}},\ \bibinfo {pages} {1330}
  (\bibinfo {year} {2004})}\BibitemShut {NoStop}%
\bibitem [{\citenamefont {Gruber}\ \emph {et~al.}(1997)\citenamefont {Gruber},
  \citenamefont {DrŠbenstedt}, \citenamefont {Tietz}, \citenamefont {Fleury},
  \citenamefont {Wrachtrup},\ and\ \citenamefont {Borczyskowski}}]{GDT}%
  \BibitemOpen
  \bibfield  {author} {\bibinfo {author} {\bibfnamefont {A.}~\bibnamefont
  {Gruber}}, \bibinfo {author} {\bibfnamefont {A.}~\bibnamefont {DrŠbenstedt}},
  \bibinfo {author} {\bibfnamefont {C.}~\bibnamefont {Tietz}}, \bibinfo
  {author} {\bibfnamefont {L.}~\bibnamefont {Fleury}}, \bibinfo {author}
  {\bibfnamefont {J.}~\bibnamefont {Wrachtrup}}, \ and\ \bibinfo {author}
  {\bibfnamefont {C.~v.}\ \bibnamefont {Borczyskowski}},\ }\href {\doibase
  10.1126/science.276.5321.2012} {\bibfield  {journal} {\bibinfo  {journal}
  {Science}\ }\textbf {\bibinfo {volume} {276}},\ \bibinfo {pages} {2012}
  (\bibinfo {year} {1997})}\BibitemShut {NoStop}%
\bibitem [{\citenamefont {Jaklevic}\ \emph {et~al.}(1964)\citenamefont
  {Jaklevic}, \citenamefont {Lambe}, \citenamefont {Silver},\ and\
  \citenamefont {Mercereau}}]{JLS}%
  \BibitemOpen
  \bibfield  {author} {\bibinfo {author} {\bibfnamefont {R.~C.}\ \bibnamefont
  {Jaklevic}}, \bibinfo {author} {\bibfnamefont {J.}~\bibnamefont {Lambe}},
  \bibinfo {author} {\bibfnamefont {A.~H.}\ \bibnamefont {Silver}}, \ and\
  \bibinfo {author} {\bibfnamefont {J.~E.}\ \bibnamefont {Mercereau}},\ }\href
  {\doibase 10.1103/PhysRevLett.12.159} {\bibfield  {journal} {\bibinfo
  {journal} {Phys. Rev. Lett.}\ }\textbf {\bibinfo {volume} {12}},\ \bibinfo
  {pages} {159} (\bibinfo {year} {1964})}\BibitemShut {NoStop}%
\bibitem [{\citenamefont {Xu}\ \emph {et~al.}(2006)\citenamefont {Xu},
  \citenamefont {Yashchuk}, \citenamefont {Donaldson}, \citenamefont
  {Rochester}, \citenamefont {Budker},\ and\ \citenamefont {Pines}}]{SVM}%
  \BibitemOpen
  \bibfield  {author} {\bibinfo {author} {\bibfnamefont {S.}~\bibnamefont
  {Xu}}, \bibinfo {author} {\bibfnamefont {V.~V.}\ \bibnamefont {Yashchuk}},
  \bibinfo {author} {\bibfnamefont {M.~H.}\ \bibnamefont {Donaldson}}, \bibinfo
  {author} {\bibfnamefont {S.~M.}\ \bibnamefont {Rochester}}, \bibinfo {author}
  {\bibfnamefont {D.}~\bibnamefont {Budker}}, \ and\ \bibinfo {author}
  {\bibfnamefont {A.}~\bibnamefont {Pines}},\ }\href@noop {} {\bibfield
  {journal} {\bibinfo  {journal} {Proc Natl Acad Sci}\ }\textbf {\bibinfo
  {volume} {103}},\ \bibinfo {pages} {12668} (\bibinfo {year}
  {2006})}\BibitemShut {NoStop}%
\bibitem [{\citenamefont {Hall}\ \emph {et~al.}(2012)\citenamefont {Hall},
  \citenamefont {Beart}, \citenamefont {Thomas}, \citenamefont {Simpson},
  \citenamefont {McGuinness}, \citenamefont {Cole}, \citenamefont {Manton},
  \citenamefont {Scholten}, \citenamefont {Jelezko}, \citenamefont {Wrachtrup},
  \citenamefont {Petrou},\ and\ \citenamefont {Hollenberg}}]{HBT}%
  \BibitemOpen
  \bibfield  {author} {\bibinfo {author} {\bibfnamefont {L.~T.}\ \bibnamefont
  {Hall}}, \bibinfo {author} {\bibfnamefont {G.~C.~G.}\ \bibnamefont {Beart}},
  \bibinfo {author} {\bibfnamefont {E.~A.}\ \bibnamefont {Thomas}}, \bibinfo
  {author} {\bibfnamefont {D.~A.}\ \bibnamefont {Simpson}}, \bibinfo {author}
  {\bibfnamefont {L.~P.}\ \bibnamefont {McGuinness}}, \bibinfo {author}
  {\bibfnamefont {J.~H.}\ \bibnamefont {Cole}}, \bibinfo {author}
  {\bibfnamefont {J.~H.}\ \bibnamefont {Manton}}, \bibinfo {author}
  {\bibfnamefont {R.~E.}\ \bibnamefont {Scholten}}, \bibinfo {author}
  {\bibfnamefont {F.}~\bibnamefont {Jelezko}}, \bibinfo {author} {\bibfnamefont
  {J.}~\bibnamefont {Wrachtrup}}, \bibinfo {author} {\bibfnamefont
  {S.}~\bibnamefont {Petrou}}, \ and\ \bibinfo {author} {\bibfnamefont
  {L.~C.~L.}\ \bibnamefont {Hollenberg}},\ }\href {\doibase 10.1038/srep00401}
  {\bibfield  {journal} {\bibinfo  {journal} {Nature Sci. Rep.}\ }\textbf
  {\bibinfo {volume} {2}} (\bibinfo {year} {2012}),\
  10.1038/srep00401}\BibitemShut {NoStop}%
\bibitem [{\citenamefont {Pham}\ \emph {et~al.}(2011)\citenamefont {Pham},
  \citenamefont {{Le Sage}}, \citenamefont {Stanwix}, \citenamefont {Yeung},
  \citenamefont {Glenn}, \citenamefont {Trifonov}, \citenamefont {Cappellaro},
  \citenamefont {Hemmer}, \citenamefont {Lukin}, \citenamefont {Park},
  \citenamefont {Yacoby},\ and\ \citenamefont {Walsworth}}]{Pham11}%
  \BibitemOpen
  \bibfield  {author} {\bibinfo {author} {\bibfnamefont {L.~M.}\ \bibnamefont
  {Pham}}, \bibinfo {author} {\bibfnamefont {D.}~\bibnamefont {{Le Sage}}},
  \bibinfo {author} {\bibfnamefont {P.~L.}\ \bibnamefont {Stanwix}}, \bibinfo
  {author} {\bibfnamefont {T.~K.}\ \bibnamefont {Yeung}}, \bibinfo {author}
  {\bibfnamefont {D.}~\bibnamefont {Glenn}}, \bibinfo {author} {\bibfnamefont
  {A.}~\bibnamefont {Trifonov}}, \bibinfo {author} {\bibfnamefont
  {P.}~\bibnamefont {Cappellaro}}, \bibinfo {author} {\bibfnamefont {P.~R.}\
  \bibnamefont {Hemmer}}, \bibinfo {author} {\bibfnamefont {M.~D.}\
  \bibnamefont {Lukin}}, \bibinfo {author} {\bibfnamefont {H.}~\bibnamefont
  {Park}}, \bibinfo {author} {\bibfnamefont {A.}~\bibnamefont {Yacoby}}, \ and\
  \bibinfo {author} {\bibfnamefont {R.~L.}\ \bibnamefont {Walsworth}},\ }\href
  {\doibase 10.1088/1367-2630/13/4/045021} {\ \textbf {\bibinfo {volume}
  {13}},\ \bibinfo {pages} {045021} (\bibinfo {year} {2011})}\BibitemShut
  {NoStop}%
\bibitem [{\citenamefont {Aleksandrov}(2010)}]{Ale2010}%
  \BibitemOpen
  \bibfield  {author} {\bibinfo {author} {\bibfnamefont {E.~B.}\ \bibnamefont
  {Aleksandrov}},\ }\href {http://stacks.iop.org/1063-7869/53/i=5/a=A04}
  {\bibfield  {journal} {\bibinfo  {journal} {Physics-Uspekhi}\ }\textbf
  {\bibinfo {volume} {53}},\ \bibinfo {pages} {487} (\bibinfo {year}
  {2010})}\BibitemShut {NoStop}%
\bibitem [{\citenamefont {Blair}(1991)}]{Bla91}%
  \BibitemOpen
  \bibfield  {author} {\bibinfo {author} {\bibfnamefont {D.~F.}\ \bibnamefont
  {Blair}},\ }\href@noop {} {\emph {\bibinfo {title} {The Detection of
  Gravitational Waves}}}\ (\bibinfo  {publisher} {Cambridge University Press},\
  \bibinfo {address} {Great Britain},\ \bibinfo {year} {1991})\BibitemShut
  {NoStop}%
\bibitem [{\citenamefont {Maze}\ \emph {et~al.}(2008)\citenamefont {Maze},
  \citenamefont {Stanwix}, \citenamefont {Hodges}, \citenamefont {Hong},
  \citenamefont {Taylor}, \citenamefont {Cappellaro}, \citenamefont {Jiang},
  \citenamefont {Dutt}, \citenamefont {Togan}, \citenamefont {Zibrov},
  \citenamefont {Yacoby}, \citenamefont {Walsworth},\ and\ \citenamefont
  {Lukin}}]{MSH}%
  \BibitemOpen
  \bibfield  {author} {\bibinfo {author} {\bibfnamefont {J.~R.}\ \bibnamefont
  {Maze}}, \bibinfo {author} {\bibfnamefont {P.~L.}\ \bibnamefont {Stanwix}},
  \bibinfo {author} {\bibfnamefont {J.~S.}\ \bibnamefont {Hodges}}, \bibinfo
  {author} {\bibfnamefont {S.}~\bibnamefont {Hong}}, \bibinfo {author}
  {\bibfnamefont {J.~M.}\ \bibnamefont {Taylor}}, \bibinfo {author}
  {\bibfnamefont {P.}~\bibnamefont {Cappellaro}}, \bibinfo {author}
  {\bibfnamefont {L.}~\bibnamefont {Jiang}}, \bibinfo {author} {\bibfnamefont
  {M.~V.~G.}\ \bibnamefont {Dutt}}, \bibinfo {author} {\bibfnamefont
  {E.}~\bibnamefont {Togan}}, \bibinfo {author} {\bibfnamefont {A.~S.}\
  \bibnamefont {Zibrov}}, \bibinfo {author} {\bibfnamefont {A.}~\bibnamefont
  {Yacoby}}, \bibinfo {author} {\bibfnamefont {R.~L.}\ \bibnamefont
  {Walsworth}}, \ and\ \bibinfo {author} {\bibfnamefont {M.~D.}\ \bibnamefont
  {Lukin}},\ }\href {\doibase 10.1038/nature07279} {\bibfield  {journal}
  {\bibinfo  {journal} {Nature}\ }\textbf {\bibinfo {volume} {455}},\ \bibinfo
  {pages} {644} (\bibinfo {year} {2008})}\BibitemShut {NoStop}%
\bibitem [{\citenamefont {Rabi}\ \emph {et~al.}(1938)\citenamefont {Rabi},
  \citenamefont {Zacharias}, \citenamefont {Millman},\ and\ \citenamefont
  {Kusch}}]{Rabi38}%
  \BibitemOpen
  \bibfield  {author} {\bibinfo {author} {\bibfnamefont {I.~I.}\ \bibnamefont
  {Rabi}}, \bibinfo {author} {\bibfnamefont {J.~R.}\ \bibnamefont {Zacharias}},
  \bibinfo {author} {\bibfnamefont {S.}~\bibnamefont {Millman}}, \ and\
  \bibinfo {author} {\bibfnamefont {P.}~\bibnamefont {Kusch}},\ }\href
  {\doibase 10.1103/PhysRev.53.318} {\bibfield  {journal} {\bibinfo  {journal}
  {Phys. Rev.}\ }\textbf {\bibinfo {volume} {53}},\ \bibinfo {pages} {318}
  (\bibinfo {year} {1938})}\BibitemShut {NoStop}%
\bibitem [{\citenamefont {Damadian}(1971)}]{Dam71}%
  \BibitemOpen
  \bibfield  {author} {\bibinfo {author} {\bibfnamefont {R.}~\bibnamefont
  {Damadian}},\ }\href {\doibase 10.1126/science.171.3976.1151} {\bibfield
  {journal} {\bibinfo  {journal} {Science}\ }\textbf {\bibinfo {volume}
  {171}},\ \bibinfo {pages} {1151} (\bibinfo {year} {1971})}\BibitemShut
  {NoStop}%
\bibitem [{\citenamefont {Bloom}(1962)}]{Bloom62}%
  \BibitemOpen
  \bibfield  {author} {\bibinfo {author} {\bibfnamefont {A.~L.}\ \bibnamefont
  {Bloom}},\ }\href {\doibase 10.1364/AO.1.000061} {\bibfield  {journal}
  {\bibinfo  {journal} {Appl. Opt.}\ }\textbf {\bibinfo {volume} {1}},\
  \bibinfo {pages} {61} (\bibinfo {year} {1962})}\BibitemShut {NoStop}%
\bibitem [{\citenamefont {Dupont-Roc}\ \emph {et~al.}(1969)\citenamefont
  {Dupont-Roc}, \citenamefont {Haroche},\ and\ \citenamefont
  {Cohen-Tannoudji}}]{DHT}%
  \BibitemOpen
  \bibfield  {author} {\bibinfo {author} {\bibfnamefont {J.}~\bibnamefont
  {Dupont-Roc}}, \bibinfo {author} {\bibfnamefont {S.}~\bibnamefont {Haroche}},
  \ and\ \bibinfo {author} {\bibfnamefont {C.}~\bibnamefont
  {Cohen-Tannoudji}},\ }\href {\doibase 10.1016/0375-9601(69)90480-0}
  {\bibfield  {journal} {\bibinfo  {journal} {Phys. Lett. A}\ }\textbf
  {\bibinfo {volume} {28}},\ \bibinfo {pages} {638 } (\bibinfo {year}
  {1969})}\BibitemShut {NoStop}%
\bibitem [{\citenamefont {Ramsey}(1950)}]{Ram50}%
  \BibitemOpen
  \bibfield  {author} {\bibinfo {author} {\bibfnamefont {N.~F.}\ \bibnamefont
  {Ramsey}},\ }\href {\doibase 10.1103/PhysRev.78.695} {\bibfield  {journal}
  {\bibinfo  {journal} {Phys. Rev.}\ }\textbf {\bibinfo {volume} {78}},\
  \bibinfo {pages} {695} (\bibinfo {year} {1950})}\BibitemShut {NoStop}%
\bibitem [{\citenamefont {Cooper}\ \emph {et~al.}(2013)\citenamefont {Cooper},
  \citenamefont {Magesan}, \citenamefont {Yum},\ and\ \citenamefont
  {Cappellaro}}]{CMYC}%
  \BibitemOpen
  \bibfield  {author} {\bibinfo {author} {\bibfnamefont {A.}~\bibnamefont
  {Cooper}}, \bibinfo {author} {\bibfnamefont {E.}~\bibnamefont {Magesan}},
  \bibinfo {author} {\bibfnamefont {H.~N.}\ \bibnamefont {Yum}}, \ and\
  \bibinfo {author} {\bibfnamefont {P.}~\bibnamefont {Cappellaro}},\
  }\href@noop {} {} (\bibinfo {year} {2013}),\ \bibinfo {note}
  {arXiv:1305.6082}\BibitemShut {NoStop}%
\bibitem [{\citenamefont {Candes}\ \emph {et~al.}(2006)\citenamefont {Candes},
  \citenamefont {Romberg},\ and\ \citenamefont {Tao}}]{CRT}%
  \BibitemOpen
  \bibfield  {author} {\bibinfo {author} {\bibfnamefont {E.~J.}\ \bibnamefont
  {Candes}}, \bibinfo {author} {\bibfnamefont {J.~K.}\ \bibnamefont {Romberg}},
  \ and\ \bibinfo {author} {\bibfnamefont {T.}~\bibnamefont {Tao}},\ }\href
  {\doibase 10.1002/cpa.20124} {\bibfield  {journal} {\bibinfo  {journal}
  {Communications on Pure and Applied Mathematics}\ }\textbf {\bibinfo {volume}
  {59}},\ \bibinfo {pages} {1207} (\bibinfo {year} {2006})}\BibitemShut
  {NoStop}%
\bibitem [{\citenamefont {Donoho}(2006)}]{Don06}%
  \BibitemOpen
  \bibfield  {author} {\bibinfo {author} {\bibfnamefont {D.}~\bibnamefont
  {Donoho}},\ }\href {\doibase 10.1109/TIT.2006.871582} {\bibfield  {journal}
  {\bibinfo  {journal} {Information Theory, IEEE Transactions on}\ }\textbf
  {\bibinfo {volume} {52}},\ \bibinfo {pages} {1289} (\bibinfo {year}
  {2006})}\BibitemShut {NoStop}%
\bibitem [{\citenamefont {Lustig}\ \emph {et~al.}(2007)\citenamefont {Lustig},
  \citenamefont {Donoho},\ and\ \citenamefont {Pauly}}]{LDP}%
  \BibitemOpen
  \bibfield  {author} {\bibinfo {author} {\bibfnamefont {M.}~\bibnamefont
  {Lustig}}, \bibinfo {author} {\bibfnamefont {D.}~\bibnamefont {Donoho}}, \
  and\ \bibinfo {author} {\bibfnamefont {J.~M.}\ \bibnamefont {Pauly}},\
  }\href@noop {} {\bibfield  {journal} {\bibinfo  {journal} {Magn. Reson.
  Med.}\ }\textbf {\bibinfo {volume} {58}},\ \bibinfo {pages} {1182} (\bibinfo
  {year} {2007})}\BibitemShut {NoStop}%
\bibitem [{\citenamefont {Taubock}\ and\ \citenamefont
  {Hlawatsch}(2008)}]{TH08}%
  \BibitemOpen
  \bibfield  {author} {\bibinfo {author} {\bibfnamefont {G.}~\bibnamefont
  {Taubock}}\ and\ \bibinfo {author} {\bibfnamefont {F.}~\bibnamefont
  {Hlawatsch}},\ }in\ \href {\doibase 10.1109/ICASSP.2008.4518252} {\emph
  {\bibinfo {booktitle} {Acoustics, Speech and Signal Processing, 2008. ICASSP
  2008. IEEE International Conference on}}}\ (\bibinfo {year} {2008})\ pp.\
  \bibinfo {pages} {2885--2888}\BibitemShut {NoStop}%
\bibitem [{\citenamefont {Dei}\ \emph {et~al.}(2009)\citenamefont {Dei},
  \citenamefont {Sheikh}, \citenamefont {Milenkovic},\ and\ \citenamefont
  {Baraniuk}}]{DSMB}%
  \BibitemOpen
  \bibfield  {author} {\bibinfo {author} {\bibfnamefont {W.}~\bibnamefont
  {Dei}}, \bibinfo {author} {\bibfnamefont {M.}~\bibnamefont {Sheikh}},
  \bibinfo {author} {\bibfnamefont {O.}~\bibnamefont {Milenkovic}}, \ and\
  \bibinfo {author} {\bibfnamefont {R.}~\bibnamefont {Baraniuk}},\ }\href@noop
  {} {\bibfield  {journal} {\bibinfo  {journal} {EURASIP J. Bioinform. Syst.
  Biol.}\ }\textbf {\bibinfo {volume} {1}},\ \bibinfo {pages} {162824}
  (\bibinfo {year} {2009})}\BibitemShut {NoStop}%
\bibitem [{\citenamefont {Lin}\ and\ \citenamefont {Herrmann}(2007)}]{LH07}%
  \BibitemOpen
  \bibfield  {author} {\bibinfo {author} {\bibfnamefont {T.~T.}\ \bibnamefont
  {Lin}}\ and\ \bibinfo {author} {\bibfnamefont {F.~J.}\ \bibnamefont
  {Herrmann}},\ }\href
  {https://www.slim.eos.ubc.ca/Publications/Public/Journals/Geophysics/2007/lin07cwe/lin07cwe.pdf}
  {\bibfield  {journal} {\bibinfo  {journal} {Geophysics}\ }\textbf {\bibinfo
  {volume} {72}},\ \bibinfo {pages} {SM77} (\bibinfo {year}
  {2007})}\BibitemShut {NoStop}%
\bibitem [{\citenamefont {Baraniuk}\ and\ \citenamefont
  {Steeghs}(2007)}]{BS07}%
  \BibitemOpen
  \bibfield  {author} {\bibinfo {author} {\bibfnamefont {R.}~\bibnamefont
  {Baraniuk}}\ and\ \bibinfo {author} {\bibfnamefont {P.}~\bibnamefont
  {Steeghs}},\ }in\ \href {\doibase 10.1109/RADAR.2007.374203} {\emph {\bibinfo
  {booktitle} {Radar Conference, 2007 IEEE}}}\ (\bibinfo {year} {2007})\ pp.\
  \bibinfo {pages} {128--133}\BibitemShut {NoStop}%
\bibitem [{\citenamefont {Gross}\ \emph {et~al.}(2010)\citenamefont {Gross},
  \citenamefont {Liu}, \citenamefont {Flammia}, \citenamefont {Becker},\ and\
  \citenamefont {Eisert}}]{GYFB}%
  \BibitemOpen
  \bibfield  {author} {\bibinfo {author} {\bibfnamefont {D.}~\bibnamefont
  {Gross}}, \bibinfo {author} {\bibfnamefont {Y.-K.}\ \bibnamefont {Liu}},
  \bibinfo {author} {\bibfnamefont {S.~T.}\ \bibnamefont {Flammia}}, \bibinfo
  {author} {\bibfnamefont {S.}~\bibnamefont {Becker}}, \ and\ \bibinfo {author}
  {\bibfnamefont {J.}~\bibnamefont {Eisert}},\ }\href {\doibase
  10.1103/PhysRevLett.105.150401} {\bibfield  {journal} {\bibinfo  {journal}
  {Phys. Rev. Lett.}\ }\textbf {\bibinfo {volume} {105}},\ \bibinfo {pages}
  {150401} (\bibinfo {year} {2010})}\BibitemShut {NoStop}%
\bibitem [{\citenamefont {Griffin}\ \emph {et~al.}(2011)\citenamefont
  {Griffin}, \citenamefont {Hirvonen}, \citenamefont {Tzagkarakis},
  \citenamefont {Mouchtaris},\ and\ \citenamefont {Tsakalides}}]{GHT11}%
  \BibitemOpen
  \bibfield  {author} {\bibinfo {author} {\bibfnamefont {A.}~\bibnamefont
  {Griffin}}, \bibinfo {author} {\bibfnamefont {T.}~\bibnamefont {Hirvonen}},
  \bibinfo {author} {\bibfnamefont {C.}~\bibnamefont {Tzagkarakis}}, \bibinfo
  {author} {\bibfnamefont {A.}~\bibnamefont {Mouchtaris}}, \ and\ \bibinfo
  {author} {\bibfnamefont {P.}~\bibnamefont {Tsakalides}},\ }\href {\doibase
  10.1109/TASL.2010.2090656} {\bibfield  {journal} {\bibinfo  {journal} {Audio,
  Speech, and Language Processing, IEEE Transactions on}\ }\textbf {\bibinfo
  {volume} {19}},\ \bibinfo {pages} {1382} (\bibinfo {year}
  {2011})}\BibitemShut {NoStop}%
\bibitem [{\citenamefont {Sen}\ and\ \citenamefont {Darabi}(2011)}]{SD11}%
  \BibitemOpen
  \bibfield  {author} {\bibinfo {author} {\bibfnamefont {P.}~\bibnamefont
  {Sen}}\ and\ \bibinfo {author} {\bibfnamefont {S.}~\bibnamefont {Darabi}},\
  }\href {\doibase http://doi.ieeecomputersociety.org/10.1109/TVCG.2010.46}
  {\bibfield  {journal} {\bibinfo  {journal} {IEEE Transactions on
  Visualization and Computer Graphics}\ }\textbf {\bibinfo {volume} {17}},\
  \bibinfo {pages} {487} (\bibinfo {year} {2011})}\BibitemShut {NoStop}%
\bibitem [{\citenamefont {Shabani}\ \emph {et~al.}(2011)\citenamefont
  {Shabani}, \citenamefont {Kosut}, \citenamefont {Mohseni}, \citenamefont
  {Rabitz}, \citenamefont {Broome}, \citenamefont {Almeida}, \citenamefont
  {Fedrizzi},\ and\ \citenamefont {White}}]{SKMR}%
  \BibitemOpen
  \bibfield  {author} {\bibinfo {author} {\bibfnamefont {A.}~\bibnamefont
  {Shabani}}, \bibinfo {author} {\bibfnamefont {R.~L.}\ \bibnamefont {Kosut}},
  \bibinfo {author} {\bibfnamefont {M.}~\bibnamefont {Mohseni}}, \bibinfo
  {author} {\bibfnamefont {H.}~\bibnamefont {Rabitz}}, \bibinfo {author}
  {\bibfnamefont {M.~A.}\ \bibnamefont {Broome}}, \bibinfo {author}
  {\bibfnamefont {M.~P.}\ \bibnamefont {Almeida}}, \bibinfo {author}
  {\bibfnamefont {A.}~\bibnamefont {Fedrizzi}}, \ and\ \bibinfo {author}
  {\bibfnamefont {A.~G.}\ \bibnamefont {White}},\ }\href {\doibase
  10.1103/PhysRevLett.106.100401} {\bibfield  {journal} {\bibinfo  {journal}
  {Phys. Rev. Lett.}\ }\textbf {\bibinfo {volume} {106}},\ \bibinfo {pages}
  {100401} (\bibinfo {year} {2011})}\BibitemShut {NoStop}%
\bibitem [{\citenamefont {Donoho}\ and\ \citenamefont {Huo}(2001)}]{DH01}%
  \BibitemOpen
  \bibfield  {author} {\bibinfo {author} {\bibfnamefont {D.}~\bibnamefont
  {Donoho}}\ and\ \bibinfo {author} {\bibfnamefont {X.}~\bibnamefont {Huo}},\
  }\href {\doibase 10.1109/18.959265} {\bibfield  {journal} {\bibinfo
  {journal} {Information Theory, IEEE Transactions on}\ }\textbf {\bibinfo
  {volume} {47}},\ \bibinfo {pages} {2845} (\bibinfo {year}
  {2001})}\BibitemShut {NoStop}%
\bibitem [{\citenamefont {Rudin}(1991)}]{Rudin91}%
  \BibitemOpen
  \bibfield  {author} {\bibinfo {author} {\bibfnamefont {W.}~\bibnamefont
  {Rudin}},\ }\href@noop {} {\emph {\bibinfo {title} {Functional Analysis}}}\
  (\bibinfo  {publisher} {McGraw-Hill Science/Engineering/Math},\ \bibinfo
  {year} {1991})\BibitemShut {NoStop}%
\bibitem [{\citenamefont {Candes}\ and\ \citenamefont {Romberg}(2007)}]{CR}%
  \BibitemOpen
  \bibfield  {author} {\bibinfo {author} {\bibfnamefont {E.}~\bibnamefont
  {Candes}}\ and\ \bibinfo {author} {\bibfnamefont {J.}~\bibnamefont
  {Romberg}},\ }\href {http://stacks.iop.org/0266-5611/23/i=3/a=008} {\bibfield
   {journal} {\bibinfo  {journal} {Inverse Problems}\ }\textbf {\bibinfo
  {volume} {23}},\ \bibinfo {pages} {969} (\bibinfo {year} {2007})}\BibitemShut
  {NoStop}%
\bibitem [{\citenamefont {Candes}\ and\ \citenamefont {Tao}(2005)}]{CT05}%
  \BibitemOpen
  \bibfield  {author} {\bibinfo {author} {\bibfnamefont {E.}~\bibnamefont
  {Candes}}\ and\ \bibinfo {author} {\bibfnamefont {T.}~\bibnamefont {Tao}},\
  }\href {\doibase 10.1109/TIT.2005.858979} {\bibfield  {journal} {\bibinfo
  {journal} {Information Theory, IEEE Transactions on}\ }\textbf {\bibinfo
  {volume} {51}},\ \bibinfo {pages} {4203} (\bibinfo {year}
  {2005})}\BibitemShut {NoStop}%
\bibitem [{\citenamefont {Foucart}(2010)}]{Foucart2010}%
  \BibitemOpen
  \bibfield  {author} {\bibinfo {author} {\bibfnamefont {S.}~\bibnamefont
  {Foucart}},\ }\href {\doibase 10.1016/j.acha.2009.10.004} {\bibfield
  {journal} {\bibinfo  {journal} {Appl. and Comp. Harmonic Analysis}\ }\textbf
  {\bibinfo {volume} {29}},\ \bibinfo {pages} {97 } (\bibinfo {year}
  {2010})}\BibitemShut {NoStop}%
\bibitem [{\citenamefont {Baraniuk}\ \emph {et~al.}(2007)\citenamefont
  {Baraniuk}, \citenamefont {Davenport}, \citenamefont {Devore},\ and\
  \citenamefont {Wakin}}]{Baraniuk2007}%
  \BibitemOpen
  \bibfield  {author} {\bibinfo {author} {\bibfnamefont {R.}~\bibnamefont
  {Baraniuk}}, \bibinfo {author} {\bibfnamefont {M.}~\bibnamefont {Davenport}},
  \bibinfo {author} {\bibfnamefont {R.}~\bibnamefont {Devore}}, \ and\ \bibinfo
  {author} {\bibfnamefont {M.}~\bibnamefont {Wakin}},\ }\href@noop {}
  {\bibfield  {journal} {\bibinfo  {journal} {Constr. Approx}\ }\textbf
  {\bibinfo {volume} {2008}} (\bibinfo {year} {2007})}\BibitemShut {NoStop}%
\bibitem [{\citenamefont {Taylor}\ \emph {et~al.}(2008)\citenamefont {Taylor},
  \citenamefont {Cappellaro}, \citenamefont {Childress}, \citenamefont {Jiang},
  \citenamefont {Budker}, \citenamefont {Hemmer}, \citenamefont {Yacoby},
  \citenamefont {Walsworth},\ and\ \citenamefont {Lukin}}]{TCC}%
  \BibitemOpen
  \bibfield  {author} {\bibinfo {author} {\bibfnamefont {J.~M.}\ \bibnamefont
  {Taylor}}, \bibinfo {author} {\bibfnamefont {P.}~\bibnamefont {Cappellaro}},
  \bibinfo {author} {\bibfnamefont {L.}~\bibnamefont {Childress}}, \bibinfo
  {author} {\bibfnamefont {L.}~\bibnamefont {Jiang}}, \bibinfo {author}
  {\bibfnamefont {D.}~\bibnamefont {Budker}}, \bibinfo {author} {\bibfnamefont
  {P.~R.}\ \bibnamefont {Hemmer}}, \bibinfo {author} {\bibfnamefont
  {A.}~\bibnamefont {Yacoby}}, \bibinfo {author} {\bibfnamefont
  {R.}~\bibnamefont {Walsworth}}, \ and\ \bibinfo {author} {\bibfnamefont
  {M.~D.}\ \bibnamefont {Lukin}},\ }\href {\doibase 10.1038/nphys1075}
  {\bibfield  {journal} {\bibinfo  {journal} {Nature Physics}\ }\textbf
  {\bibinfo {volume} {4}},\ \bibinfo {pages} {2417} (\bibinfo {year}
  {2008})}\BibitemShut {NoStop}%
\bibitem [{\citenamefont {Hahn}(1950)}]{Hahn50}%
  \BibitemOpen
  \bibfield  {author} {\bibinfo {author} {\bibfnamefont {E.~L.}\ \bibnamefont
  {Hahn}},\ }\href {\doibase 10.1103/PhysRev.80.580} {\bibfield  {journal}
  {\bibinfo  {journal} {Phys. Rev.}\ }\textbf {\bibinfo {volume} {80}},\
  \bibinfo {pages} {580} (\bibinfo {year} {1950})}\BibitemShut {NoStop}%
\bibitem [{\citenamefont {Carr}\ and\ \citenamefont {Purcell}(1954)}]{CP}%
  \BibitemOpen
  \bibfield  {author} {\bibinfo {author} {\bibfnamefont {H.~Y.}\ \bibnamefont
  {Carr}}\ and\ \bibinfo {author} {\bibfnamefont {E.~M.}\ \bibnamefont
  {Purcell}},\ }\href {\doibase 10.1103/PhysRev.94.630} {\bibfield  {journal}
  {\bibinfo  {journal} {Phys. Rev.}\ }\textbf {\bibinfo {volume} {94}},\
  \bibinfo {pages} {630} (\bibinfo {year} {1954})}\BibitemShut {NoStop}%
\bibitem [{\citenamefont {Meiboom}\ and\ \citenamefont {Gill}(1958)}]{MG}%
  \BibitemOpen
  \bibfield  {author} {\bibinfo {author} {\bibfnamefont {S.}~\bibnamefont
  {Meiboom}}\ and\ \bibinfo {author} {\bibfnamefont {D.}~\bibnamefont {Gill}},\
  }\href@noop {} {\bibfield  {journal} {\bibinfo  {journal} {Rev. Sci.
  Instrum.}\ }\textbf {\bibinfo {volume} {29}},\ \bibinfo {pages} {688}
  (\bibinfo {year} {1958})}\BibitemShut {NoStop}%
\bibitem [{\citenamefont {Walsh}(1923)}]{Walsh23}%
  \BibitemOpen
  \bibfield  {author} {\bibinfo {author} {\bibfnamefont {J.~L.}\ \bibnamefont
  {Walsh}},\ }\href@noop {} {\bibfield  {journal} {\bibinfo  {journal} {Amer.
  J. Math..}\ }\textbf {\bibinfo {volume} {45}},\ \bibinfo {pages} {5}
  (\bibinfo {year} {1923})}\BibitemShut {NoStop}%
\bibitem [{\citenamefont {Dayan}\ and\ \citenamefont {Abbott}(2005)}]{DA}%
  \BibitemOpen
  \bibfield  {author} {\bibinfo {author} {\bibfnamefont {P.}~\bibnamefont
  {Dayan}}\ and\ \bibinfo {author} {\bibfnamefont {L.}~\bibnamefont {Abbott}},\
  }\href@noop {} {\emph {\bibinfo {title} {Theoretical Neuroscience:
  Computational and Mathematical Modeling of Neural Systems}}}\ (\bibinfo
  {publisher} {MIT Press},\ \bibinfo {address} {U.S.A.},\ \bibinfo {year}
  {2005})\BibitemShut {NoStop}%
\bibitem [{\citenamefont {Hayes}\ \emph {et~al.}(2011)\citenamefont {Hayes},
  \citenamefont {Khodjasteh}, \citenamefont {Viola},\ and\ \citenamefont
  {Biercuk}}]{HKV}%
  \BibitemOpen
  \bibfield  {author} {\bibinfo {author} {\bibfnamefont {D.}~\bibnamefont
  {Hayes}}, \bibinfo {author} {\bibfnamefont {K.}~\bibnamefont {Khodjasteh}},
  \bibinfo {author} {\bibfnamefont {L.}~\bibnamefont {Viola}}, \ and\ \bibinfo
  {author} {\bibfnamefont {M.~J.}\ \bibnamefont {Biercuk}},\ }\href {\doibase
  10.1103/PhysRevA.84.062323} {\bibfield  {journal} {\bibinfo  {journal} {Phys.
  Rev. A}\ }\textbf {\bibinfo {volume} {84}},\ \bibinfo {pages} {062323}
  (\bibinfo {year} {2011})}\BibitemShut {NoStop}%
\bibitem [{\citenamefont {Spivak}(1967)}]{Spivak}%
  \BibitemOpen
  \bibfield  {author} {\bibinfo {author} {\bibfnamefont {M.}~\bibnamefont
  {Spivak}},\ }\href@noop {} {\emph {\bibinfo {title} {Calculus}}}\ (\bibinfo
  {publisher} {Cambridge University Press},\ \bibinfo {address} {Cambridge,
  UK},\ \bibinfo {year} {1967})\BibitemShut {NoStop}%
\bibitem [{\citenamefont {Woosley}\ \emph {et~al.}(1985)\citenamefont
  {Woosley}, \citenamefont {Roth},\ and\ \citenamefont {Jr.}}]{WRW}%
  \BibitemOpen
  \bibfield  {author} {\bibinfo {author} {\bibfnamefont {J.~K.}\ \bibnamefont
  {Woosley}}, \bibinfo {author} {\bibfnamefont {B.~J.}\ \bibnamefont {Roth}}, \
  and\ \bibinfo {author} {\bibfnamefont {J.~P.~W.}\ \bibnamefont {Jr.}},\
  }\href {\doibase 10.1016/0025-5564(85)90044-6} {\bibfield  {journal}
  {\bibinfo  {journal} {Math. Biosciences}\ }\textbf {\bibinfo {volume} {76}},\
  \bibinfo {pages} {1 } (\bibinfo {year} {1985})}\BibitemShut {NoStop}%
\bibitem [{\citenamefont {Chow}\ \emph {et~al.}(2008)\citenamefont {Chow},
  \citenamefont {Dagens}, \citenamefont {Fu}, \citenamefont {Cook},\ and\
  \citenamefont {Paley}}]{CDF}%
  \BibitemOpen
  \bibfield  {author} {\bibinfo {author} {\bibfnamefont {L.~S.}\ \bibnamefont
  {Chow}}, \bibinfo {author} {\bibfnamefont {A.}~\bibnamefont {Dagens}},
  \bibinfo {author} {\bibfnamefont {Y.}~\bibnamefont {Fu}}, \bibinfo {author}
  {\bibfnamefont {G.~G.}\ \bibnamefont {Cook}}, \ and\ \bibinfo {author}
  {\bibfnamefont {M.~N.}\ \bibnamefont {Paley}},\ }\href {\doibase
  10.1002/mrm.21753} {\bibfield  {journal} {\bibinfo  {journal} {Magnetic
  Resonance in Medicine}\ }\textbf {\bibinfo {volume} {60}},\ \bibinfo {pages}
  {1147} (\bibinfo {year} {2008})}\BibitemShut {NoStop}%
\bibitem [{\citenamefont {Anwar}\ \emph {et~al.}(2009)\citenamefont {Anwar},
  \citenamefont {Cook}, \citenamefont {Chow},\ and\ \citenamefont
  {Paley}}]{ACCP}%
  \BibitemOpen
  \bibfield  {author} {\bibinfo {author} {\bibfnamefont {S.~M.}\ \bibnamefont
  {Anwar}}, \bibinfo {author} {\bibfnamefont {G.~G.}\ \bibnamefont {Cook}},
  \bibinfo {author} {\bibfnamefont {L.~S.}\ \bibnamefont {Chow}}, \ and\
  \bibinfo {author} {\bibfnamefont {M.~N.}\ \bibnamefont {Paley}},\ }in\
  \href@noop {} {\emph {\bibinfo {booktitle} {Proc. Intl. Soc. Mag. Reson. Med.
  17}}}\ (\bibinfo {year} {2009})\BibitemShut {NoStop}%
\bibitem [{\citenamefont {Murakami}\ and\ \citenamefont {Okada}(2006)}]{MO}%
  \BibitemOpen
  \bibfield  {author} {\bibinfo {author} {\bibfnamefont {S.}~\bibnamefont
  {Murakami}}\ and\ \bibinfo {author} {\bibfnamefont {Y.}~\bibnamefont
  {Okada}},\ }\href {\doibase 10.1113/jphysiol.2006.105379} {\bibfield
  {journal} {\bibinfo  {journal} {The Journal of Physiology}\ }\textbf
  {\bibinfo {volume} {575}},\ \bibinfo {pages} {925} (\bibinfo {year}
  {2006})}\BibitemShut {NoStop}%
\bibitem [{\citenamefont {Uhrig}(2008)}]{Uhrig08}%
  \BibitemOpen
  \bibfield  {author} {\bibinfo {author} {\bibfnamefont {G.}~\bibnamefont
  {Uhrig}},\ }\href {http://stacks.iop.org/1367-2630/10/i=8/a=083024}
  {\bibfield  {journal} {\bibinfo  {journal} {New J. Phys.}\ }\textbf {\bibinfo
  {volume} {10}},\ \bibinfo {pages} {083024} (\bibinfo {year}
  {2008})}\BibitemShut {NoStop}%
\bibitem [{\citenamefont {Cywi\ifmmode~\acute{n}\else \'{n}\fi{}ski}\ \emph
  {et~al.}(2008)\citenamefont {Cywi\ifmmode~\acute{n}\else \'{n}\fi{}ski},
  \citenamefont {Lutchyn}, \citenamefont {Nave},\ and\ \citenamefont
  {Das~Sarma}}]{CLL}%
  \BibitemOpen
  \bibfield  {author} {\bibinfo {author} {\bibfnamefont {L.}~\bibnamefont
  {Cywi\ifmmode~\acute{n}\else \'{n}\fi{}ski}}, \bibinfo {author}
  {\bibfnamefont {R.~M.}\ \bibnamefont {Lutchyn}}, \bibinfo {author}
  {\bibfnamefont {C.~P.}\ \bibnamefont {Nave}}, \ and\ \bibinfo {author}
  {\bibfnamefont {S.}~\bibnamefont {Das~Sarma}},\ }\href {\doibase
  10.1103/PhysRevB.77.174509} {\bibfield  {journal} {\bibinfo  {journal} {Phys.
  Rev. B}\ }\textbf {\bibinfo {volume} {77}},\ \bibinfo {pages} {174509}
  (\bibinfo {year} {2008})}\BibitemShut {NoStop}%
\bibitem [{\citenamefont {Bylander}\ \emph {et~al.}(2011)\citenamefont
  {Bylander}, \citenamefont {Gustavsson}, \citenamefont {Yan}, \citenamefont
  {Yoshihara}, \citenamefont {Harrabi}, \citenamefont {Fitch}, \citenamefont
  {Cory}, \citenamefont {Yasunobu}, \citenamefont {Shen},\ and\ \citenamefont
  {Oliver}}]{BGY}%
  \BibitemOpen
  \bibfield  {author} {\bibinfo {author} {\bibfnamefont {J.}~\bibnamefont
  {Bylander}}, \bibinfo {author} {\bibfnamefont {S.}~\bibnamefont
  {Gustavsson}}, \bibinfo {author} {\bibfnamefont {F.}~\bibnamefont {Yan}},
  \bibinfo {author} {\bibfnamefont {F.}~\bibnamefont {Yoshihara}}, \bibinfo
  {author} {\bibfnamefont {K.}~\bibnamefont {Harrabi}}, \bibinfo {author}
  {\bibfnamefont {G.}~\bibnamefont {Fitch}}, \bibinfo {author} {\bibfnamefont
  {D.}~\bibnamefont {Cory}}, \bibinfo {author} {\bibfnamefont {N.}~\bibnamefont
  {Yasunobu}}, \bibinfo {author} {\bibfnamefont {J.-S.}\ \bibnamefont {Shen}},
  \ and\ \bibinfo {author} {\bibfnamefont {W.}~\bibnamefont {Oliver}},\ }\href
  {\doibase 10.1038/nphys1994} {\bibfield  {journal} {\bibinfo  {journal}
  {Nature Physics}\ }\textbf {\bibinfo {volume} {7}},\ \bibinfo {pages} {565}
  (\bibinfo {year} {2011})}\BibitemShut {NoStop}%
\bibitem [{\citenamefont {Bar-Gill}\ \emph {et~al.}(2012)\citenamefont
  {Bar-Gill}, \citenamefont {Pham}, \citenamefont {Belthangady}, \citenamefont
  {Le~Sage}, \citenamefont {Cappellaro}, \citenamefont {Maze}, \citenamefont
  {Lukin}, \citenamefont {Yacoby},\ and\ \citenamefont {Walsworth}}]{BPB}%
  \BibitemOpen
  \bibfield  {author} {\bibinfo {author} {\bibfnamefont {N.}~\bibnamefont
  {Bar-Gill}}, \bibinfo {author} {\bibfnamefont {L.}~\bibnamefont {Pham}},
  \bibinfo {author} {\bibfnamefont {C.}~\bibnamefont {Belthangady}}, \bibinfo
  {author} {\bibfnamefont {D.}~\bibnamefont {Le~Sage}}, \bibinfo {author}
  {\bibfnamefont {P.}~\bibnamefont {Cappellaro}}, \bibinfo {author}
  {\bibfnamefont {J.}~\bibnamefont {Maze}}, \bibinfo {author} {\bibfnamefont
  {M.}~\bibnamefont {Lukin}}, \bibinfo {author} {\bibfnamefont
  {A.}~\bibnamefont {Yacoby}}, \ and\ \bibinfo {author} {\bibfnamefont
  {R.}~\bibnamefont {Walsworth}},\ }\href {\doibase 10.1038/ncomms1856}
  {\bibfield  {journal} {\bibinfo  {journal} {Nat. Commun.}\ }\textbf {\bibinfo
  {volume} {2}},\ \bibinfo {pages} {858} (\bibinfo {year} {2012})}\BibitemShut
  {NoStop}%
\bibitem [{\citenamefont {Magesan}\ \emph {et~al.}(2013)\citenamefont
  {Magesan}, \citenamefont {Cooper}, \citenamefont {Yum},\ and\ \citenamefont
  {Cappellaro}}]{MCYC}%
  \BibitemOpen
  \bibfield  {author} {\bibinfo {author} {\bibfnamefont {E.}~\bibnamefont
  {Magesan}}, \bibinfo {author} {\bibfnamefont {A.}~\bibnamefont {Cooper}},
  \bibinfo {author} {\bibfnamefont {H.~N.}\ \bibnamefont {Yum}}, \ and\
  \bibinfo {author} {\bibfnamefont {P.}~\bibnamefont {Cappellaro}},\
  }\href@noop {} {} (\bibinfo {year} {2013}),\ \bibinfo {note}
  {arXiv:1305.6604}\BibitemShut {NoStop}%
\bibitem [{\citenamefont {Paley}(1932)}]{Paley32}%
  \BibitemOpen
  \bibfield  {author} {\bibinfo {author} {\bibfnamefont {R.~E. A.~C.}\
  \bibnamefont {Paley}},\ }\href@noop {} {\bibfield  {journal} {\bibinfo
  {journal} {Proc. London Math. Soc.}\ }\textbf {\bibinfo {volume} {34}},\
  \bibinfo {pages} {241} (\bibinfo {year} {1932})}\BibitemShut {NoStop}%
\bibitem [{\citenamefont {Gray}(1953)}]{Gray53}%
  \BibitemOpen
  \bibfield  {author} {\bibinfo {author} {\bibfnamefont {F.}~\bibnamefont
  {Gray}},\ }\href@noop {} {\bibfield  {journal} {\bibinfo  {journal} {U.S.
  patent no. 2632058}\ } (\bibinfo {year} {1953})}\BibitemShut {NoStop}%
\end{thebibliography}

%

\end{document}